\documentclass[11pt,prd,aps,floats,preprint,eqsecnum,nofootinbib]{revtex4}
\usepackage[hypertex]{hyperref}
\usepackage{graphicx,amsmath,amssymb,verbatim,mathrsfs,array,layout,textcomp,latexsym}
\usepackage{multirow}
\usepackage{color}

\newcommand{\be}{\begin{equation}}
\newcommand{\ee}{\end{equation}}
\newcommand{\bea}{\begin{eqnarray}}
\newcommand{\eea}{\end{eqnarray}}
\newcommand{\beq}{\begin{equation}}
\newcommand{\eeq}{\end{equation}}
\newcommand{\beqa}{\begin{eqnarray}}
\newcommand{\eeqa}{\end{eqnarray}}
\newcommand{\no}{\nonumber}

\newcommand{\cL}{\mathcal{L}}
\newcommand{\cO}{\mathcal{O}}
\newcommand{\cA}{\mathcal{A}}
\newcommand{\cB}{\mathcal{B}}

\newcommand{\Qbar}{\overline{Q}}
\newcommand{\Dbar}{\overline{D}}
\newcommand{\Ebar}{\overline{E}}
\newcommand{\Lbar}{\overline{L}}
\newcommand{\Bbar}{\overline{B}}
\newcommand{\Mbar}{\overline{M}}
\newcommand{\Knotbar}{\overline{K^0}}
\newcommand{\Dnotbar}{\overline{D^0}}

\newcommand{\Nkk}{{N_{\rm KK}}}
\newcommand{\Mkk}{{M_{\rm KK}}}
\newcommand{\Mkks}{{M^2_{\rm KK}}}
\newcommand{\yfd}{{y_{\rm 5D}}}
\newcommand{\yfdmin}{{y_{\rm 5D}^{\rm min}}}

\newcommand{\rg}{{r^g_{00}}}

\definecolor{red}{cmyk}{0,1,1,0.4}

\def\lsim{\mathrel{\rlap{\lower4pt\hbox{\hskip1pt$\sim$}}
     \raise1pt\hbox{$<$}}}         
\def\gsim{\mathrel{\rlap{\lower4pt\hbox{\hskip1pt$\sim$}}
     \raise1pt\hbox{$>$}}}         

\begin{document}

%
\title{Flavor Physics Constraints\\ for Physics Beyond the Standard Model}
\author{Gino Isidori\footnote{Email: gino.isidori@lnf.infn.it}}
\affiliation{INFN, Laboratori Nazionali di Frascati, 00044 Frascati,
Italy,\\ and TUM~Institite~for Advanced Study, 80333 M\"unchen, Germany}
\author{Yosef Nir\footnote{The Amos de-Shalit chair of theoretical
    physics}\footnote{Email: yosef.nir@weizmann.ac.il}
and Gilad Perez\footnote{Email: gilad.perez@weizmann.ac.il}}
   \affiliation{Department of Particle Physics and Astrophysics,
   Weizmann Institute of Science, Rehovot 76100, Israel\vspace*{1cm}}

\vspace*{1cm}
\begin{abstract}
In the last decade, huge progress in experimentally measuring and
theoretically understanding flavor physics has been achieved. In
particular, the accuracy in the determination of the CKM elements has
been greatly improved, and a large number of flavor changing neutral
current processes, involving $b\to d$, $b\to s$ and $c\to u$
transitions, and of CP violating asymmetries, have been measured. No
evidence for new physics has been established. Consequently, strong
constraints on new physics at high scale apply. In particular, the
flavor structure of new physics at the TeV scale is strongly
constrained. We review these constraints  and we discuss
the future prospects to better understand the flavor structure
of physics beyond the Standard Model.
\end{abstract}
\maketitle


\section{Introduction}
\label{sec:intro}
The term ``{\bf flavors}'' is used, in the jargon of particle physics, to
describe several copies of the same gauge representation, namely
several fields that are assigned the same quantum charges. Within the
Standard Model (SM), when thinking of its unbroken $SU(3)_{\rm C}\times U(1)_{\rm
  EM}$ gauge group, there are four different types of particles, each
coming in three flavors:
\begin{itemize}
\item Up-type quarks in the $(3)_{+2/3}$ representation: $u,c,t$;
\item Down-type quarks in the $(3)_{-1/3}$ representation: $d,s,b$;
\item Charged leptons in the $(1)_{-1}$ representation: $e,\mu,\tau$;
\item Neutrinos in the $(1)_{0}$ representation: $\nu_1,\nu_2,\nu_3$.
\end{itemize}

The term ``{\bf flavor physics}'' refers to interactions that distinguish
between flavors. By definition, gauge interactions, namely
interactions that are related to unbroken symmetries and mediated
therefore by massless gauge bosons, do not distinguish among the
flavors and do not constitute part of flavor physics. Within the
Standard Model, flavor-physics refers to the weak and Yukawa
interactions. With New Physics (NP), there are likely to be additional
`flavored' interactions.

The term ``{\bf flavor changing}'' refers to processes where the
initial and final flavor-numbers (that is, the number of particles of a
certain flavor minus the number of anti-particles of the same flavor)
are different. In ``flavor changing charged current'' processes, both
up-type and down-type flavors, and/or both charged lepton and neutrino
flavors are involved. Examples are (i) muon decay via $\mu\to
e\bar\nu_i\nu_j$, and (ii) $K^-\to\mu^-\bar\nu_j$ (which corresponds,
at the quark level, to $s\bar u\to\mu^-\bar\nu_j$). Within the
Standard Model, these processes are mediated by the $W$-bosons and
occur at tree level. In ``{\bf flavor changing neutral current}'' (FCNC)
processes, either up-type or down-type flavors but not both, and/or
either charged lepton or neutrino flavors but not both, are involved.
Example are (i) muon decay via $\mu\to e\gamma$ and (ii)
$K_L\to\mu^+\mu^-$ (which corresponds, at the quark level, to $s\bar
d\to\mu^+\mu^-$). Within the Standard Model, these processes do not
occur at tree level, and are often highly suppressed. This situation
makes FCNC particularly sensitive to new physics: If the new physics
does not have the same flavor suppression factors as the standard
model, then it could contribute to FCNC comparably to the standard
model even if it takes places at energy scales that are orders of
magnitude higher than the weak scale. The fact that flavor physics is
a very sensitive probe of high energy physics is the main reason for
the experimental effort to measure flavor parameters and the
theoretical effort to interpret these data.

\section{The Standard Model}
\label{sec:sm}
The Standard Model (SM) is defined as follows:\\
(i) The gauge symmetry is
\beq\label{smsym}
G_{\rm SM}=SU(3)_{\rm C}\times SU(2)_{\rm L}\times U(1)_{\rm Y}.
\eeq
It is spontaneously broken by the
vacuum expectation values (VEV) of a single Higgs scalar,
$\phi(1,2)_{1/2}$ ($\langle\phi^0\rangle=v/\sqrt{2}$):
\beq\label{smssb}
G_{\rm SM} \to SU(3)_{\rm C}\times U(1)_{\rm EM}.
\eeq
(ii) There are three fermion generations, each consisting of five
representations of $G_{\rm SM}$:
\beq\label{ferrep}
Q_{Li}(3,2)_{+1/6},\ \ U_{Ri}(3,1)_{+2/3},\ \
D_{Ri}(3,1)_{-1/3},\ \ L_{Li}(1,2)_{-1/2},\ \ E_{Ri}(1,1)_{-1}.
\eeq
The Standard Model Lagrangian, ${\cal L}_{\rm SM}$, is the most
general renormalizable Lagrangian that is consistent with the gauge
symmetry (\ref{smsym}), the particle content (\ref{ferrep}) and the
pattern of spontaneous symmetry breaking (\ref{smssb}). It can be
divided to three parts:
\beq\label{LagSM}
{\cal L}_{\rm SM}={\cal L}_{\rm kinetic+gauge}+{\cal L}_{\rm Higgs}
+{\cal L}_{\rm Yukawa}.
\eeq

The source of all flavor physics is in the Yukawa interactions:
\beq\label{Hqint}
-{\cal L}_{\rm Yukawa}=Y^d_{ij}~{\Qbar_{Li}}\phi D_{Rj}
+Y^u_{ij}~{\Qbar_{Li}}\tilde\phi U_{Rj}+Y^e_{ij}~{\Lbar_{Li}}\phi
E_{Rj} +{\rm h.c.}.
\eeq
(where $\tilde\phi=i\tau_2\phi^\dagger$).
This part of the Lagrangian is, in general, flavor-dependent (that is,
$Y^f\not\propto{\bf 1}$) and CP violating.

In the absence of the Yukawa matrices $Y^d$, $Y^u$ and $Y^e$, the SM
has a large $U(3)^5$ global symmetry:
\beq\label{gglobal}
G_{\rm global}(Y^{u,d,e}=0)=SU(3)_q^3\times SU(3)_\ell^2\times U(1)^5.
\eeq
The non-Abelian part of this symmetry is particularly relevant to
flavor physics:
\beqa\label{susuu}
SU(3)_q^3&=&SU(3)_Q\times SU(3)_U\times SU(3)_D,\no\\
SU(3)_\ell^2&=&SU(3)_L\times SU(3)_E.
\eeqa
The point that is important for our purposes is that ${\cal L}_{\rm
  kinetic+gauge}+{\cal L}_{\rm Higgs}$ respect the non-Abelian flavor
symmetry $S(3)_q^3\times SU(3)_\ell^2$, under which
\beq\label{symkh}
Q_L\to V_QQ_L,\ \ \ U_R\to V_U U_R,\ \ \ D_R\to V_D D_R,\ \ L_L\to V_L
L_L,\ \ \ E_R\to V_E E_R,
\eeq
where the $V_i$ are unitary matrices.
The Yukawa interactions (\ref{Hqint}) break the global symmetry,
\beq\label{globre}
G_{\rm global}(Y^{u,d,e}\neq0)= U(1)_B\times U(1)_e\times
U(1)_\mu\times U(1)_\tau.
\eeq
(Of course, the gauged $U(1)_Y$ also remains a good symmetry.)
Thus, the transformations of Eq. (\ref{symkh}) are not a symmetry of
${\cal L}_{\rm SM}$. Instead, they correspond to a change of the
interaction basis. These observations also offer an alternative way of
defining flavor physics: it refers to interactions that break the
$SU(3)^5$ symmetry (\ref{symkh}). Thus, the term ``{\bf flavor
  violation}'' is often used to describe processes or parameters that
break the symmetry.

Using the transformation (\ref{symkh}), one can choose an interaction
basis where the number of parameters is minimized. A useful example of
such a basis for the quark Yukawa matrices is the following:
\beq\label{speint}
Y^d=\lambda_d,\ \ \ Y^u=V^\dagger\lambda_u,
\eeq
where $\lambda_{d,u}$ are diagonal,
\beq\label{deflamd}
\lambda_d={\rm diag}(y_d,y_s,y_b),\ \ \
\lambda_u={\rm diag}(y_u,y_c,y_t),
\eeq
while $V$ is a unitary matrix that depends on three real angles and
one complex phase. We conclude that there are 10 quark flavor
parameters: 9 real ones and a single phase. In the mass basis, one
identifies six of the real parameters as the six quark masses, while
the remaining three real and one imaginary parameters appear in the
CKM matrix \cite{Cabibbo:1963yz,Kobayashi:1973fv}, $V$, which
describes the couplings of the charged weak-force carriers, the
$W^\pm$-bosons, with quark-antiquark pairs.

Within the standard model, all flavor changing physics comes from the
$V$ matrix. There are various ways to choose the four parameters of
$V$. One of the most convenient ways is given by the Wolfenstein
parametrization, where the four mixing parameters are
$(\lambda,A,\rho,\eta)$ with $\lambda=|V_{us}|=0.23$ playing the role
of an expansion parameter and $\eta$ representing the CP violating
phase \cite{Wolfenstein:1983yz,Buras:1994ec}:
\beq\label{wolpar}
V\simeq\left(\begin{matrix}
1-\frac12\lambda^2-\frac18\lambda^4 & \lambda &
A\lambda^3(\rho-i\eta)\cr
-\lambda +\frac12A^2\lambda^5[1-2(\rho+i\eta)] &
1-\frac12\lambda^2-\frac18\lambda^4(1+4A^2) & A\lambda^2 \cr
A\lambda^3[1-(1-\frac12\lambda^2)(\rho+i\eta)]&
-A\lambda^2+\frac12A\lambda^4[1-2(\rho+i\eta)]
& 1-\frac12A^2\lambda^4 \cr\end{matrix}\right) .
\eeq

\begin{figure}[tb]
  \centering
  {\includegraphics[width=0.65\textwidth]{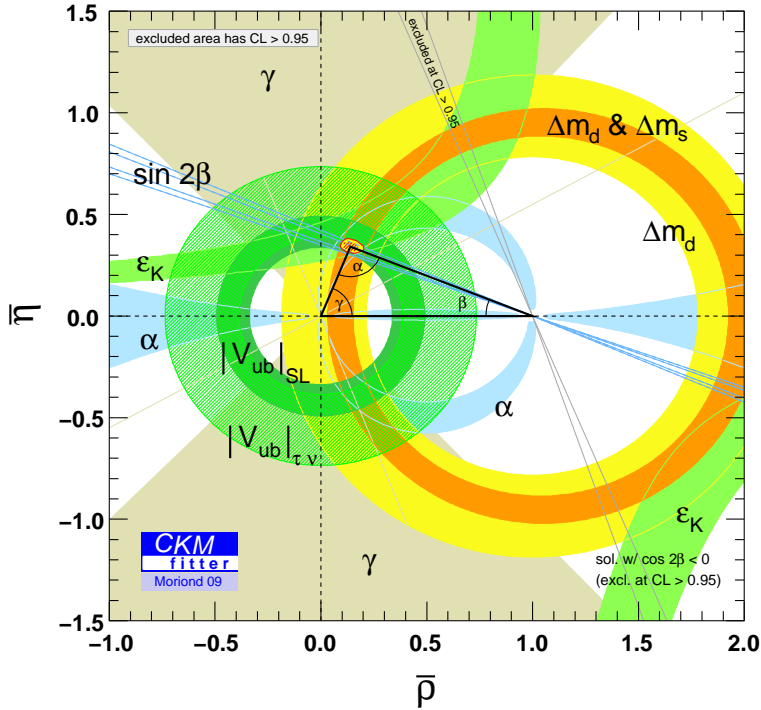}}
  \caption{Allowed region in the $\rho,\eta$ plane. Superimposed are
  the individual constraints from charmless semileptonic $B$ decays
  ($|V_{ub}/V_{cb}|$), mass differences in the $B^0$ ($\Delta m_d$)
  and $B_s$ ($\Delta m_s$) neutral meson systems, and CP violation in
  $K\to\pi\pi$ ($\varepsilon_K$), $B\to\psi K$ ($\sin2\beta$),
  $B\to\pi\pi,\rho\pi,\rho\rho$ ($\alpha$), and $B\to DK$
  ($\gamma$). Taken from \cite{ckmfitter}.}
  \label{fg:UT}
\end{figure}

One can assume that flavor changing processes are
fully described by the SM, and check the consistency of the various
measurements with this assumption. The
values of $\lambda$ and $A$ are known rather accurately
\cite{Amsler:2008zzb} from, respectively, $K\to\pi\ell\nu$ and $b\to
c\ell\nu$ decays:
\beq\label{lamaexp}
\lambda=0.2257\pm0.0010,\ \ \ A=0.814\pm0.022.
\eeq
Then, one can express all the relevant observables as a function of
the two remaining parameters, $\rho$ and $\eta$, and check whether
there is a range in the $\rho-\eta$ plane that is consistent with all
measurements. The list of observables includes the following:
\begin{itemize}
\item The rates of inclusive and exclusive charmless semileptonic $B$
  decays depend on $|V_{ub}|^2\propto\rho^2+\eta^2$;
\item The CP asymmetry in $B\to\psi K_S$, $S_{B\to \psi
    K}=\sin2\beta=\frac{2\eta(1-\rho)}{(1-\rho)^2+\eta^2}$;
\item The rates of various $B\to DK$ decays depend on the phase
  $\gamma$, where $e^{i\gamma}=\frac{\rho+i\eta}{\sqrt{\rho^2+\eta^2}}$;
\item The rates of various $B\to\pi\pi,\rho\pi,\rho\rho$ decays depend
  on the phase $\alpha=\pi-\beta-\gamma$;
\item The ratio between the mass splittings in the neutral $B$ and
  $B_s$ systems is sensitive to
  $|V_{td}/V_{ts}|^2=\lambda^2[(1-\rho)^2+\eta^2]$;
\item The CP violation in $K\to\pi\pi$ decays, $\epsilon_K$, depends
  in a complicated way on $\rho$ and $\eta$.
\end{itemize}
The resulting constraints are shown in Fig. \ref{fg:UT}.

The consistency of the various constraints is impressive. In
particular, the following ranges for $\rho$ and $\eta$ can account for
all the measurements \cite{Amsler:2008zzb}:
\beq
\rho=+0.135^{+0.031}_{-0.016},\ \ \ \eta=+0.349\pm0.017.
\eeq

One can make then the following statement \cite{Nir:2002gu,Ligeti:2004ak,NMFV}:\\
{\bf Very likely, flavor violation and CP violation in flavor changing
  processes are dominated by the CKM mechanism.}

One can actually go a step further, and allow for arbitrary new
physics in all flavor changing processes except for those that have
contributions from SM tree diagrams. Then, one can quantitatively
constrain the size of new physics contributions to processes such as
neutral meson mixing. We do so in the next section.

\section{Model Independent Constraints}
\label{sec:indep}
In order to describe NP effects in flavor physics we can follow two
main strategies: (i) build an explicit ultraviolet completion of the
model, and specify which are the new fields beyond the SM ones, or (ii)
analyse the NP effects using a generic effective-theory approach, or
integrating-out the new heavy fields.  The first approach is more
predictive, but also more model dependent. We follow this approach in
Sect.~\ref{sec:susy} and~\ref{sec:exdim} in two well-motivated SM
extensions.  In this and the next section we follow the second
strategy, which is less predictive but also more general.

Assuming the new degrees to be heavier than SM fields, we can
integrate them out and describe NP effects by means of a
generalization of the Fermi Theory.  The SM Lagrangian becomes the
renormalizable part of a more general local Lagrangian which includes
an infinite tower of operators with dimension $d>4$, constructed in
terms of SM fields, suppressed by inverse powers of an effective scale
$\Lambda > M_W$:
\be
\cL_{\rm eff} =
\cL_{\rm SM}  + \sum ~ \frac{c_{i}^{(d)}}{\Lambda^{(d-4)}} ~
O_i^{(d)}({\rm SM~fields}).
\label{eq:effL}
\ee
This general bottom-up approach allows us to analyse all realistic
extensions of the SM in terms of a limited number of parameters (the
coefficients of the higher-dimensional operators).  The drawback of
this method is the impossibility to establish correlations of NP
effects at low and high energies: the scale $\Lambda$ defines the
cut-off of the effective theory.  However, correlations among
different low-energy processes can still be established implementing
specific symmetry properties, such as the MFV hypothesis
(Sect.~\ref{sec:mfv}).  The experimental tests of such correlations
allow us to test/establish general features of the new theory which
holds independently of the dynamical details of the model.  In
particular, $B$, $D$ and $K$ decays are extremely useful in
determining the flavor-symmetry breaking pattern of the NP model.

\subsection{Bounds from $\Delta F=2$ down-type transitions}
The starting points for this analysis is the observation that in
several realistic NP models we can neglect non-standard effects in all
cases where the corresponding effective operator is generated at the
tree-level within the SM. This general assumption implies that the
experimental determination of the CKM matrix via tree-level processes
is free from the contamination of NP contributions.  Using this
determination we can unambiguously predict meson-antimeson mixing and
FCNC amplitudes within the SM and compare it with data, constraining
the couplings of the $\Delta F=2$ operators in~(\ref{eq:effL}). Each
$\Delta F=2$ amplitude is then conveniently parametrized in terms of
the shift induced in the modulo and the CPV phase, or the real and
imaginary part~\cite{Silva:1996ih,Grossman:1997dd}:
\begin{equation}
\frac{\langle B_q| \cL_\mathrm{eff}| \Bbar_q\rangle}{\langle
    B_q| \cL_\mathrm{SM} | \Bbar_q\rangle}
    =C_{B_q}  e^{2 i \phi_{B_q}}~, \qquad
\frac{{\rm Re}[\langle K^0| \cL_\mathrm{eff}| \Knotbar \rangle]}{{\rm Re}[\langle
    K^0 | \cL_\mathrm{SM} | \Knotbar \rangle]}
    =C_{\Delta m_K} ~\stackrel{{\rm Re} \to {\rm Im}}{\longrightarrow}~
     C_{\epsilon_K}~,
\end{equation}
An updated analysis of these constraints has been presented
in~\cite{Bona:2007vi} (see Fig.~\ref{fig:UTfit}). The main conclusions
that can be drawn from this analysis can be summarized as follows:
\begin{figure}[t]
\begin{center}
\includegraphics[width=70mm]{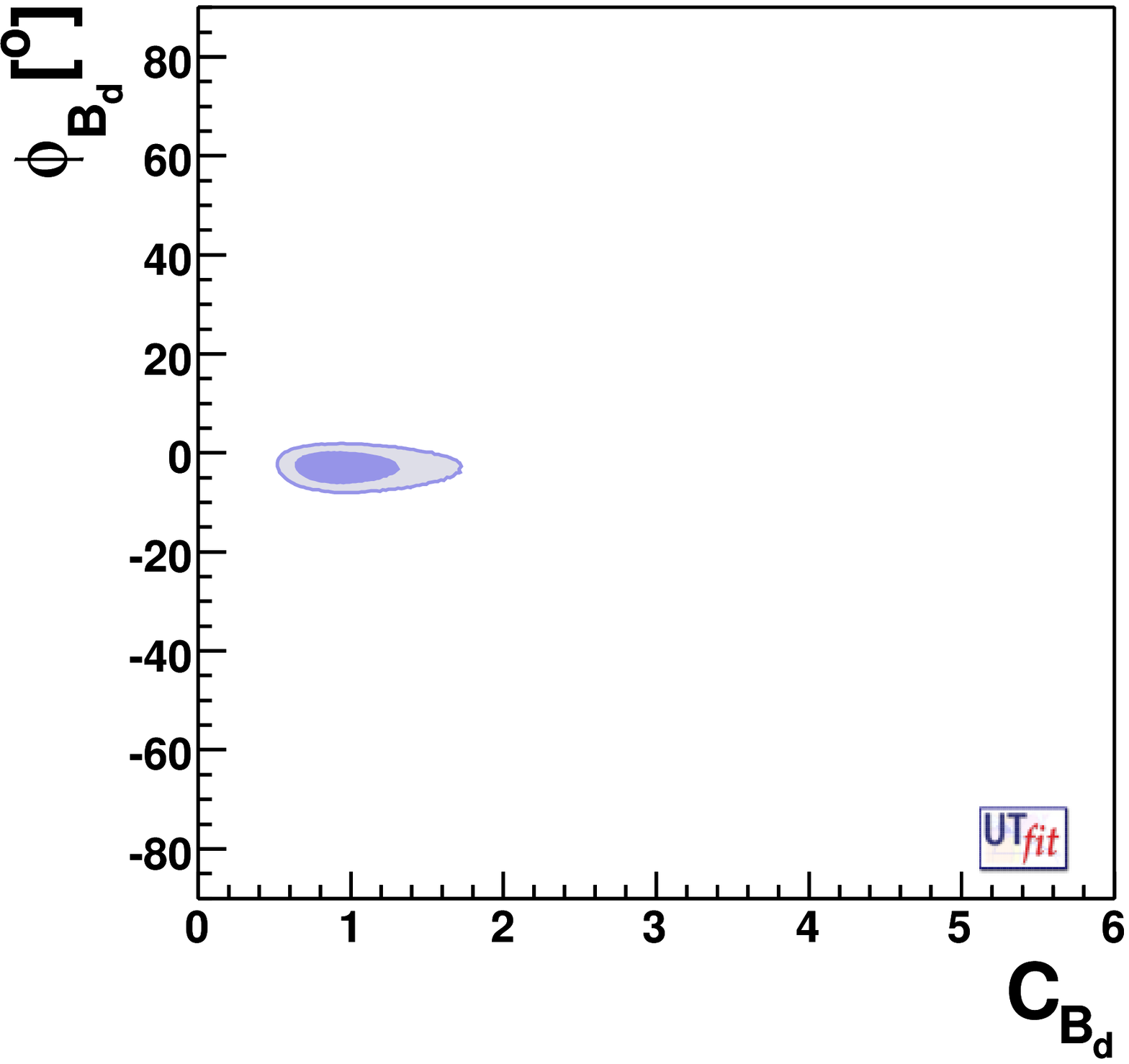}
\includegraphics[width=70mm]{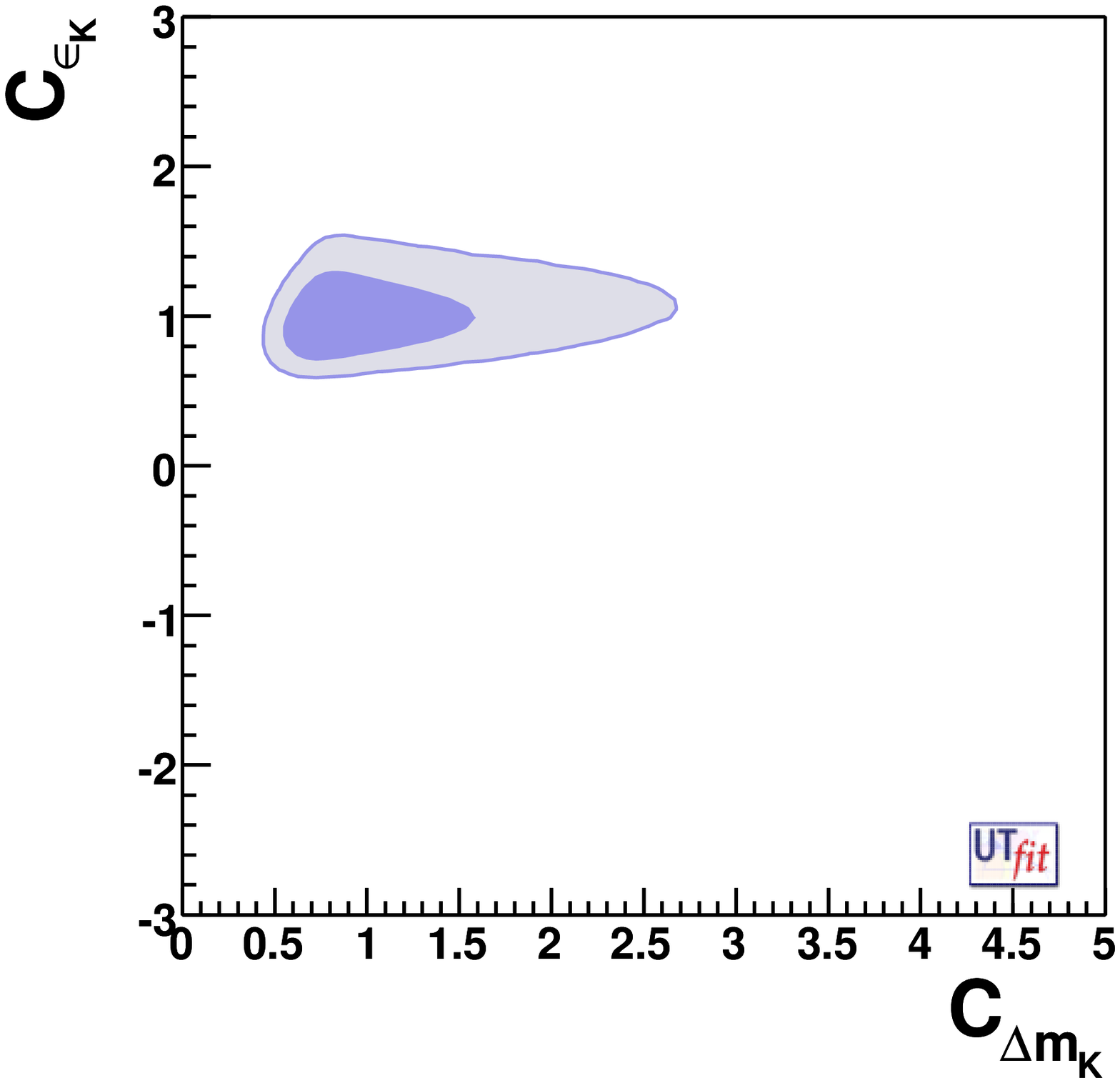}
\caption{%
  Constraints on the effective parameters encoding NP effects
  in $B_d$--$\Bbar_d$ mixing and $K^0$--$\Knotbar$ mixing as
  obtained by the UTfit collaboration~\cite{Bona:2007vi}.}
\label{fig:UTfit}
\end{center}
\end{figure}

(i) In all the three accessible short-distance amplitudes
($K^0$--$\Knotbar$, $B_d$--$\Bbar_d$, and $B_s$--$\Bbar_s$) the
magnitude of the new-physics amplitude cannot exceed, in size, the SM
short-distance contribution.  The latter is suppressed by both the GIM
mechanism and the hierarchical structure of the CKM matrix,
\be
 \cA_{\rm SM}^{\Delta F=2} \approx
   \frac{ G_F^2 m_t^2 }{16 \pi^2} \left(V_{ti}^* V_{tj} \right)^2
\times
   \langle \Mbar |  (\Qbar_{Li} \gamma^\mu Q_{Lj} )^2  | M \rangle
\times    F\left(\frac{M_W^2}{m_t^2}\right),
\ee
where $F$ is a loop function of $\cO(1)$. As a result, new-physics
models with TeV-scale flavored degrees of freedom and $\mathcal{O}(1)$
effective flavor-mixing couplings are ruled out. To set explicit
bounds, let us consider for instance the subset of left-handed $\Delta
F=2$ operators in the generic effective Lagrangian in (\ref{eq:effL}),
namely
\begin{equation}\label{eq:qlql}
\Delta \mathcal{L}^{\Delta F=2} = \sum_{i\not=j} \frac{c_{ij}}{\Lambda^2}
(\Qbar_{Li} \gamma^\mu Q_{Lj} )^2~,
\end{equation}
where  the $c_{ij}$ are dimensionless couplings. The condition
$|\mathcal{A}^{\Delta F=2}_{\rm NP}| <  |\mathcal{A}^{\Delta F=2}_{\rm SM} |$
implies
\begin{eqnarray}
\Lambda > \frac{ 4.4~{\rm TeV} }{| V_{ti}^* V_{tj}|/|c_{ij}|^{1/2}  }
\sim \left\{ \begin{array}{l}
1.3\times 10^4~{\rm TeV} \times |c_{sd}|^{1/2} \!\!\!\!\!\!\! \\
5.1\times 10^2~{\rm TeV} \times |c_{bd}|^{1/2} \!\!\!\!\!\!\! \\
1.1\times 10^2~{\rm TeV} \times |c_{bs}|^{1/2} \!\!\!\!\!\!\!
\end{array}
\right.
\label{eq:bound}
\end{eqnarray}
The strong bounds on $\Lambda$ for generic $c_{ij}$ of order 1 is a
manifestation of what in many specific frameworks (supersymmetry,
technicolor, etc.)  goes under the name of {\em flavor problem}: if we
insist that the new physics emerges in the TeV region, we have to
conclude that it possesses a highly non-generic flavor structure.

(ii) In the case of $B_d$--$\Bbar_d$ and $K^0$--$\Knotbar$ mixing,
where both CP conserving and CP-violating observables are measured
with excellent accuracy, there is still room for a sizable NP
contribution (relative to the SM one), provided that it is to a good
extent aligned in phase with the SM amplitude [${\cal
  O}\left(0.01\right)$ for the $K$ system and ${\cal
  O}\left(0.3\right)$ for the $B_d$ system].  This is because the
theoretical errors in the observables used to constraint the phases,
$S_{B_d \to \psi K}$ and $\epsilon_K$, are smaller with respect to the
theoretical uncertainties in $\Delta m_{B_d}$ and $\Delta m_K$, which
constrain the magnitude of the mixing amplitudes.

(iii) In the case of $B_s$--$\Bbar_s$ mixing, the precise
determination of $\Delta m_{B_s}$ does not allow large deviations in
modulo with respect to the SM. The constraint is particularly severe
if we consider the ratio $\Delta m_{B_d}/\Delta m_{B_s}$, where
hadronic uncertainties cancel to a large extent. However, the
constraint on the CP-violating phase is quite poor. Present data from
CDF~\cite{Aaltonen:2007he} and D0~\cite{:2008fj} indicate a large
central value for the CP-violating phase, contrary to the SM
expectation. The errors are, however, still large and the disagreement
with the SM is at about the $2 \sigma$ level.  If the disagreement
persists, becoming statistically significant, this would not only
signal the presence of physics beyond the SM, but would also rule out
a whole subclass of MFV models (see Sect.~\ref{sec:mfv}).

(iv) In $D-\bar D$ mixing we cannot estimate
the SM contribution from first principles; however,
to a good accuracy this is CP conserving. As a result,
strong bounds on possible non-standard CP-violating
contributions can still be set.
The resulting constraints are only second to those from $\epsilon_K$,
and unlike in the case of $\epsilon_K$ are controlled by experimental
statistics and could possibly be significantly improved in the near future.

\begin{table}[t]
\begin{center}
\begin{tabular}{c|c c|c c|c} \hline\hline
\rule{0pt}{1.2em}%
Operator &  \multicolumn{2}{c|}{Bounds on $\Lambda$~in~TeV~($c_{ij}=1$)} &
\multicolumn{2}{c|}{Bounds on
$c_{ij}$~($\Lambda=1$~TeV) }& Observables\cr
&   Re& Im & Re & Im \cr
 \hline $(\bar s_L \gamma^\mu d_L )^2$  &~$9.8 \times 10^{2}$& $1.6 \times 10^{4}$
&$9.0 \times 10^{-7}$& $3.4 \times 10^{-9}$ & $\Delta m_K$; $\epsilon_K$ \\
($\bar s_R\, d_L)(\bar s_L d_R$)   & $1.8 \times 10^{4}$& $3.2 \times 10^{5}$
&$6.9 \times 10^{-9}$& $2.6 \times 10^{-11}$ &  $\Delta m_K$; $\epsilon_K$ \\
 \hline $(\bar c_L \gamma^\mu u_L )^2$  &$1.2 \times 10^{3}$& $2.9 \times 10^{3}$
&$5.6 \times 10^{-7}$& $1.0 \times 10^{-7}$ & $\Delta m_D$; $|q/p|, \phi_D$ \\
($\bar c_R\, u_L)(\bar c_L u_R$)   & $6.2 \times 10^{3}$& $1.5 \times 10^{4}$
&$5.7 \times 10^{-8}$& $1.1 \times 10^{-8}$ &  $\Delta m_D$; $|q/p|, \phi_D$\\
\hline$(\bar b_L \gamma^\mu d_L )^2$    &  $5.1 \times 10^{2}$ & $9.3
\times 10^{2}$ &  $3.3 \times 10^{-6}$ &
$1.0 \times 10^{-6}$ & $\Delta m_{B_d}$; $S_{\psi K_S}$  \\
($\bar b_R\, d_L)(\bar b_L d_R)$  &   $1.9 \times 10^{3}$ & $3.6
\times 10^{3}$ &  $5.6 \times 10^{-7}$ &   $1.7 \times 10^{-7}$
&   $\Delta m_{B_d}$; $S_{\psi K_S}$ \\
\hline $(\bar b_L \gamma^\mu s_L )^2$    &  \multicolumn{2}{c|}{$1.1 \times 10^{2}$} &
 \multicolumn{2}{c|}{$7.6\times10^{-5}$}  & $\Delta m_{B_s}$ \\
($\bar b_R \,s_L)(\bar b_L s_R)$  &   \multicolumn{2}{c|}{$3.7 \times 10^{2}$}   &
 \multicolumn{2}{c|}{$1.3\times10^{-5}$} & $\Delta m_{B_s}$ \\ \hline\hline
\end{tabular}
\caption{\label{tab:DF2} Bounds on representative dimension-six
  $\Delta F=2$
operators. Bounds on $\Lambda$ are quoted
assuming an effective coupling $1/\Lambda^2$, or, alternatively, the bounds on the respective $c_{ij}$'s assuming
$\Lambda=1$ TeV. Observables related to CPV are
separated from the CP conserving ones with semicolons.
In the $B_s$ system we only quote a bound on the modulo of the NP amplitude
derived from $\Delta m_{B_s}$  (see text).
For the definition of the CPV observables in the $D$ system see
Ref.~\cite{Bergmann:2000id}. }
\end{center}
\end{table}

A more detailed list of the bounds derived from $\Delta F=2$
observables is shown in Table~\ref{tab:DF2}, where we quote the bounds
for two representative sets of dimension-six operators: the left-left
operators (present also in the SM) and operators with a different
chirality, which arise in specific SM extensions. The bounds on the
latter are stronger, especially in the kaon case, because of the larger
hadronic matrix elements.  The
constraints related to CPV correspond to maximal phases, and are
subject to the requirement that the NP contributions are smaller than
$30\%$ ($60\%$) of the total contributions~\cite{NMFV} in the $B_d$ ($K$)
system.  Since the experimental status of CP violation in the $B_s$ system is not yet settled we simply require
that the new physics contributions are smaller than the observed value of $\Delta m_{B_s}$ (for less naive treatments see {\it e.g.}~\cite{Bona:2007vi,ckmfitter}).

\subsection{Correlations between $K$ and $D$ mixing}
There are two different features that can provide flavor-related
suppression factors: degeneracy and alignment. In general, low energy
measurements can only constrain the product of these two suppression
factors. An interesting exception occurs, however, for the LL
operators of the type (\ref{eq:qlql}) where there is an independent
constraint on the level of degeneracy \cite{Blum:2009sk}. We here
briefly explain this point.

Consider operators of the form
\beq\label{xqoperator}
\frac{1}{\Lambda_{\rm NP}^2}(\Qbar_{Li}(X_Q)_{ij}\gamma_\mu
Q_{Lj}) (\Qbar_{Li}(X_Q)_{ij}\gamma^\mu Q_{Lj}),
\eeq
where $X_Q$ is an hermitian matrix. Without loss of
generality, we can choose to work in the basis defined in
Eq. (\ref{speint}):
\beq
Y^d=\lambda_d,\ \ \ Y^u=V^\dagger\lambda_u,\ \ \
X_Q=V_d^\dagger\lambda_Q V_d,
\eeq
where $\lambda_{Q}$ is a diagonal real matrix, and $V_d$ is a
unitary matrix which parametrizes the misalignment of the operator
(\ref{xqoperator}) with the down mass basis.

The experimental constraints that are most relevant to our study come
from $K^0$--$\Knotbar$ and $D^0$--$\Dnotbar$ mixing, which involve
only the first two generation quarks. When studying new physics
effects, ignoring the third generation is often a good approximation
to the physics at hand. Indeed, even when the third generation does
play a role, our two generation analysis is applicable as long as
there are no strong cancellations with contributions related to the
third generation.  In a two generation framework, $V$ depends on a
single mixing angle (the Cabibbo angle $\theta_c$), while $V_d$
depends on a single angle and a single phase. To understand various
aspects of our analysis, it is useful, however, to provisionally set
the phase to zero, and study only CP conserving (CPC) observables. We
thus have
\beqa\label{twogen}
\lambda_Q&=&{\rm diag}(\lambda_1,\lambda_2),\ \ \
V=\left(\begin{array}{cc}
  \cos\theta_c & \sin\theta_c \\ -\sin\theta_c & \cos\theta_c
\end{array}\right),\ \ \
V_d=\left(\begin{array}{cc}
    \cos\theta_d & \sin\theta_d \\ -\sin\theta_d & \cos\theta_d
  \end{array}\right).
\eeqa
It is convenient to define
\beq
\lambda_{12}=\frac12(\lambda_{1}+\lambda_{2}),\ \
\delta_{12}=\frac{\lambda_{1}-\lambda_{2}}{\lambda_{1}+\lambda_{2}},\ \
\Lambda_{12}=\delta_{12}\lambda_{12}.
\eeq
Thus $\lambda_{12}$ parametrizes the overall, flavor-diagonal
suppression of $X_Q$ (in particular, loop factors),
$\delta_{12}$ parametrizes suppression that is coming from approximate
degeneracy between the eigenvalues of $X_Q$, and $\theta_d$ and
$\theta_c-\theta_d$ parametrize the suppression that comes from
alignment with, respectively, the down and the up sector.

The main point is the following: Alignment can entirely suppress the
contribution to either $K^0$--$\Knotbar$ mixing ($\theta_d=0$) or
$D^0$--$\Dnotbar$ mixing ($\theta_d=\theta_c$) but not to both.
Thus, the flavor measurements give a constraint on $\Lambda_{12}$ which
reads \cite{Blum:2009sk}
\beq\label{boucpc}
\Lambda_{12}\leq3.8\times10^{-3}
\left(\frac{\Lambda_{\rm NP}}{1\ {\rm TeV}}\right).
\eeq
If we switch on the CP violating phase $\gamma$ in $V_d$ then, for
$0.03\lsim|\sin\gamma|\lsim0.98$, we find  \cite{Blum:2009sk}
\beq\label{boucpv}
\Lambda_{12}\leq\frac{4.8\times10^{-4}}{\sqrt{\sin2\gamma}}
\left(\frac{\Lambda_{\rm NP}}{1\ {\rm TeV}}\right).
\eeq
We learn that, with a loop suppression of order
$\lambda_{12}\sim\alpha_2$, the degeneracy should be stronger than
$0.02$.

\subsection{Top Physics}
While the present direct determination of top-quark decay rates is at
the ${\cal O}(1\%)$ level of accuracy~\cite{TevT}, orders of magnitude
improvement is expected at the LHC.  With $100\,{\rm fb}^{-1}$ of
data, the LHC will be sensitive (at 95\% CL) to branching ratios of
${\cal O}\left(10^{-5}\right)$ in the $t\to u^i\,Z/\gamma$
channels~\cite{Carvalho:2007yi}, where $u^i=u,c$. Within the SM, these
channels have branching ratios in the $10^{-13}$ range, so any
experimental observation would be a clear sign of physics beyond the
SM.  The LHC sensitivity to additional $\Delta t=1$ processes, such as
$t\to c\, G,\phi$, is more limited (see {\it
  e.g.}~Ref.~\cite{tfcncelse} and references therein).

The impact of the projected LHC bounds on top flavor violation can be
described in a model independent manner. The leading contributions to
the above processes are mediated via a set of dimension-six
operators~\cite{tEFT}, which can be classified according to the SM
global flavor symmetries~\cite{Fox:2007in}. Focusing on the chirality
of the quark fields, these can be written as ${\cal O}_{A^t B}$, where
$A^t,B=L,R$ and $A^t$ denotes top-quark chirality. Bounds from $B$
physics can give severe constraints on the operators with
$B=L$~\cite{Fox:2007in}. The situation changes, however, in case of a
flavor alignment of these effective operators, such that they are
diagonal in the down-type mass basis. In this limit no bounds can be
derived on the ${\cal O}_{LR}$ and ${\cal O}_{RL}$
operators~\cite{GMP}.  On the other hand, the ${\cal O}_{L L}$
operators cannot be aligned simultaneously with the down- and up-quark
mass bases. As a result, combining bounds on $t\to u^i$ with bounds
from $b\to s l^+ l^-$ we can get stringent constraints on the operator
${\cal O}^h_{LL}=\bar Q_i \gamma^\mu
\left(X_Q\right)_{ij}Q_j\left(\phi^\dagger \overleftrightarrow{D}_\mu
  \phi\right)/\Lambda^2_t$. Normalising the $SU(3)_Q$ adjoint spurion,
$X_Q= a_i T_i$, such that the projection onto the $SU(3)$ generators
has a unit norm ($\sum_i a_i^2 =1$), we can indentify a unique
combination which minimizes the combined bound from top and $B$
constraints.  Taking into account the projected LHC sensitivity on
$t\to u^i$ decays, we can expect to reach bounds of $\cO(600\rm~GeV)$
on the scale $\Lambda_t$~\cite{GMP}, in the absence of a NP signal.

\section{Minimal Flavor Violation}
\label{sec:mfv}
A very reasonable, although quite pessimistic, set up which avoids the
new physics flavor problem is the so-called Minimal Flavor Violation
(MFV) hypothesis.  Under this assumption, flavor-violating
interactions are linked to the known structure of Yukawa couplings
also beyond the SM.  As a result, non-standard contributions in FCNC
transitions turn out to be suppressed to a level consistent with
experiments even for $\Lambda \sim$~few TeV.  One of the most
interesting aspects of the MFV hypothesis is that it can naturally be
implemented within the generic effective Lagrangian in
Eq.~(\ref{eq:effL}). Furthermore, SM extensions where the flavor
hierarchy is generated at a scale much higher than other dynamical
scales tend to flow to the MFV class of models in the infra-red.

The MFV hypothesis consists of two
ingredients~\cite{D'Ambrosio:2002ex}: (i)~a {\em flavor symmetry} and
(ii)~a set of {\em symmetry-breaking terms}.  The symmetry is nothing
but the large global symmetry of the SM Lagrangian in absence of
Yukawa couplings shown in Eq.~(\ref{susuu}). Since this global
symmetry, and particularly the ${\rm SU}(3)$ subgroups controlling
quark flavor-changing transitions, is already broken within the SM, we
cannot promote it to be an exact symmetry of the NP model. Some
breaking would appear at the quantum level because of the SM Yukawa
interactions.  The most restrictive assumption we can make to {\em
  protect} in a consistent way quark-flavor mixing beyond the SM is to
assume that $Y^d$ and $Y^u$ are the only sources of flavor symmetry
breaking also in the NP model.  To implement and interpret this
hypothesis in a consistent way, we can assume that $SU(3)^3_{q}$ is a
good symmetry and promote $Y^{u,d}$ to be non-dynamical fields
(spurions) with non-trivial transformation properties under
$SU(3)^3_{q}$:
\begin{equation}
Y^u \sim (3, \bar 3, 1)~,\qquad
Y^d \sim (3, 1, \bar 3)~.\qquad
\end{equation}
If the breaking of the symmetry occurs at very high energy scales, at
low-energies we would only be sensitive to the background values of
the $Y$, {\it i.e.}~to the ordinary SM Yukawa couplings.  Employing the
effective-theory language, an effective theory satisfies the criterion
of Minimal Flavor Violation in the quark sector\footnote{~The notion of MFV
can be extended also to the letpon sector. However, in this case
there is not a unique way to define the minimal sources of
flavour symmetry breaking if we want to keep track of non-vanishing
neutrino masses~\cite{Cirigliano:2005ck,Davidson:2006bd}.} if all
higher-dimensional operators, constructed from SM and $Y$ fields, are
invariant  (formally) under the flavor group
$SU(3)^3_{q}$~\cite{D'Ambrosio:2002ex}.
Invariance under CP may or may not be imposed in addition.

According to this criterion one should in principle consider operators
with arbitrary powers of the (dimensionless) Yukawa fields. However,
a strong simplification arises by the observation that all the
eigenvalues of the Yukawa matrices are small,
but for the top one (and possibly the bottom one, see later), and
that the off-diagonal elements of the CKM matrix are very suppressed.
Working in the basis in Eq.~(\ref{speint}), and neglecting
the ratio of light quark masses over the top mass,
we have
\be
\left[  Y^u (Y^u)^\dagger \right]^n_{i\not = j} ~\approx~
y_t^{2n} V^*_{ti} V_{tj}~.
\label{eq:basicspurion}
\ee
As a consequence, including high powers
of the the Yukawa matrices amounts only to a redefinition
of the overall factor in (\ref{eq:basicspurion}) and the
the leading $\Delta F=2$ and $\Delta F=1$ FCNC amplitudes get exactly
the same CKM suppression as in the SM:
\begin{eqnarray}
  \mathcal{A}(d^i \to d^j)_{\rm MFV} &=&   (V^*_{ti} V_{tj})^{\phantom{a}}
 \mathcal{A}^{(\Delta F=1)}_{\rm SM}
\left[ 1 + a_1 \frac{ 16 \pi^2 M^2_W }{ \Lambda^2 } \right]~,
\\
  \mathcal{A}(M_{ij}-{\Mbar_{ij}})_{\rm MFV}  &=&  (V^*_{ti} V_{tj})^2
 \mathcal{A}^{(\Delta F=2)}_{\rm SM}
\left[ 1 + a_2 \frac{ 16 \pi^2 M^2_W }{ \Lambda^2 } \right]~,
\label{eq:FC}
\end{eqnarray}
where the $\mathcal{A}^{(i)}_{\rm SM}$ are the SM loop amplitudes
and the $a_i$ are $\mathcal{O}(1)$ real parameters. The  $a_i$
depend on the specific operator considered but are flavor
independent. This implies the same relative correction
in $s\to d$, $b\to d$, and  $b\to s$ transitions
of the same type.

Within the MFV framework, several of the constraints used to determine
the CKM matrix (and in particular the unitarity triangle) are not
affected by NP~\cite{Buras:2000dm}.  In this framework, NP effects are
negligible not only in tree-level processes but also in a few clean
observables sensitive to loop effects, such as the time-dependent CPV
asymmetry in $B_d \to \psi K_{L,S}$. Indeed the structure of the basic
flavor-changing coupling in Eq.~(\ref{eq:FC}) implies that the weak
CPV phase of $B_d$--$\Bbar_d$ mixing is arg[$(V_{td}V_{tb}^*)^2$],
exactly as in the SM.  This construction provides a natural (a
posteriori) justification of why no NP effects have been observed in
the quark sector: by construction, most of the clean observables
measured at $B$ factories are insensitive to NP effects in the MFV
framework.

\begin{table}[t]
\begin{minipage}{\textwidth}
\begin{center}
\begin{tabular}{l|c|l}
Operator & ~Bound on $\Lambda$~  & ~Observables \\
\hline\hline
$H^\dagger \left( \Dbar_R Y^{d\dagger}  Y^u Y^{u\dagger}
  \sigma_{\mu\nu} Q_L \right) (e F_{\mu\nu})$
& ~$6.1$~TeV & ~$B\to X_s \gamma$, $B\to X_s \ell^+ \ell^-$\\
$\frac{1}{2} (\Qbar_L  Y^u Y^{u\dagger} \gamma_{\mu} Q_L)^2
\phantom{X^{X^X}_{iii}}$
& ~$5.9$~TeV & ~$\epsilon_K$, $\Delta m_{B_d}$, $\Delta m_{B_s}$ \\
$H_D^\dagger \left( \Dbar_R  Y^{d\dagger}
Y^u Y^{u\dagger}  \sigma_{\mu\nu}  T^a  Q_L \right) (g_s G^a_{\mu\nu})$
&~$3.4$~TeV & ~$B\to X_s \gamma$, $B\to X_s \ell^+ \ell^-$\\
$\left( \Qbar_L Y^u Y^{u\dagger}  \gamma_\mu
Q_L \right) (\Ebar_R \gamma_\mu E_R)$
& ~$2.7$~TeV & ~$B\to X_s \ell^+ \ell^-$, $B_s\to\mu^+\mu^-$ \\
$~i \left( \Qbar_L Y^u Y^{u\dagger}  \gamma_\mu
Q_L \right) H_U^\dagger D_\mu H_U$
&~$2.3$~TeV 
&~$B\to X_s \ell^+ \ell^-$, $B_s\to\mu^+\mu^-$\\
$\left( \Qbar_L Y^u Y^{u\dagger}  \gamma_\mu Q_L \right)
( \Lbar_L \gamma_\mu L_L)$
&~$1.7$~TeV & ~$B\to X_s \ell^+ \ell^-$, $B_s\to\mu^+\mu^-$\\
$\left( \Qbar_L Y^u Y^{u\dagger}  \gamma_\mu Q_L
\right) (e D_\mu F_{\mu\nu})$
&~$1.5$~TeV & ~$B\to X_s \ell^+ \ell^-$\\
\end{tabular}
\end{center}
\end{minipage}
\caption{\label{tab:MFV} Bounds on the scale of new physics (at 95\%
  C.L.) for some representative $\Delta F=1$~\cite{Hurth:2008jc} and
$\Delta F=2$~\cite{Bona:2007vi} MFV operators
(assuming effective coupling $\pm 1/\Lambda^2$), and corresponding
observables used to set the bounds.}
\end{table}

In Table~\ref{tab:MFV} we report a few representative examples of the
bounds on the higher-dimensional operators in the MFV framework.  For
simplicity, only leading spurion dependence is shown on the
left-handed column.  The built-in CKM suppression leads to bounds on
the effective scale of new physics not far from the TeV region. These
bounds are very similar to the bounds on flavor-conserving operators
derived by precision electroweak tests.  This observation reinforces
the conclusion that a deeper study of rare decays is definitely needed
in order to clarify the flavor problem: the experimental precision on
the clean FCNC observables required to obtain bounds more stringent
than those derived from precision electroweak tests (and possibly
discover new physics) is typically in the $1\%-10\%$ range.

Although MFV seems to be a natural solution to the flavor problem, it
should be stressed that (i) this is not a theory of flavor (there is
no explanation for the observed hierarchical structure of the
Yukawas), and (ii) we are still far from having proved the validity of this
hypothesis from data (in the effective theory language we can say that
there is still room for sizable new sources of flavor symmetry
breaking beside the SM Yukawa couplings~\cite{Feldmann:2006jk}).  A
proof of the MFV hypothesis can be achieved only with a positive
evidence of physics beyond the SM exhibiting the flavor-universality
pattern (same relative correction in $s\to d$, $b\to d$, and $b\to s$
transitions of the same type) predicted by the MFV assumption.  While
this goal is quite difficult to be achieved, the MFV framework is
quite predictive and thus could easily be falsified: in
Table~\ref{tab:MFVbounds} we list some clean MFV predictions which
could be falsified by future experiments.
Violations of these bounds would not only imply physics
beyond the SM, but also a clear signal of new sources of flavor
symmetry breaking beyond the Yukawa couplings.

\begin{table}[t]
\begin{minipage}{\textwidth}
\begin{center}
\begin{tabular}{l|l|l|l}
 Observable & ~Experiment  & ~{MFV prediction}~ & ~{SM prediction} \\
  \hline
  \hline
  $ \beta_s$~from~$\cA_{\rm CP}(B_s \to \psi \phi)$ & ~[0.10, 1.44] @ 95\% CL~
  &  ~$0.04(5)^*$  & ~$0.04(2)$   \\  \hline
  $\cA_{\rm CP}(B \to X_s \gamma)$ &  ~$< 6\%$ @ 95\% CL~
  &  ~$<0.02^*$  & ~$<0.01$   \\  \hline
  $\cB(B_d \to \mu^+ \mu^-)$ & ~$<1.8 \times 10^{-8}$
  \qquad    & ~$<1.2 \times 10^{-9}$ & ~$1.3(3)\times 10^{-10}$ \\
  \hline
  $\cB(B \to X_s \tau^+ \tau^-)$ & ~ -- ~  &
  ~$< 5 \times 10^{-7}$ & ~$1.6(5)\times 10^{-7}$ \\
 \hline
  $\cB(K_L \to \pi^0 \nu \bar \nu)$ & ~$<2.6 \times 10^{-8}$ @ 90\% CL
  & ~$<2.9\times 10^{-10}$ & ~$2.9(5)\times 10^{-11}$ \\
\end{tabular}
\end{center}
\end{minipage}
\caption{Some predictions derived in the MFV framework.
Stars implies that the prediction is not valid in the GMFV case at
large $\tan\beta$ (see Section~\ref{GMFV}).
\label{tab:MFVbounds} }
\end{table}

The idea that the CKM matrix rules the strength of FCNC transitions
also beyond the SM has become a very popular concept in recent
literature and has been implemented and discussed in several works.
It is worth stressing that the CKM matrix represents only one part of
the problem: a key role in determining the structure of FCNCs is also
played by quark masses, or by the Yukawa eigenvalues. In this respect,
the MFV criterion provides the maximal protection of FCNCs (or the
minimal violation of flavor symmetry), since the full structure of
Yukawa matrices is preserved. At the same time, this criterion is
based on a renormalization-group-invariant symmetry argument, which
can be implemented independently of any specific hypothesis about the
dynamics of the new-physics framework.
This model-independent structure
does not hold in most of the alternative definitions of MFV models
that can be found in the literature. For instance, the definition of
Ref.~\cite{Buras:2003jf} (denoted constrained MFV, or CMFV) contains
the additional requirement that the effective FCNC operators playing a
significant role within the SM are the only relevant ones also beyond
the SM.  This condition is realized only in weakly coupled theories at
the TeV scale with only one light Higgs doublet, such as the MSSM with
small $\tan\beta$ where no large logs, or sizable anomalous dimension
are present~\cite{Kagan:2009bn} (see also \cite{Feldmann:2008ja}).
It does not hold in several other frameworks,
such as Higgsless models, 5D MFV models, or the MSSM with large $\tan\beta$.

In NP models where sizable anomalous dimensions are present, the
expansion in powers of the Yukawa spurions cannot be truncated at the
first non-trivial order. In this limit, denoted as General MFV (GMFV),
higher-order terms in the third-generation Yukawa couplings need to be
re-summed. As shown in Eq.~(\ref{eq:basicspurion}), if only the
top-quark Yukawa coupling is of order one, this resummation has
negligible consequences for flavour-violating processes. (To measure
the effect requires that the accuracy on rare $K$ ($D$) decays would
become strong enough to probe the contributions related to the charm
(strange) quark Yukawa coupling~\cite{Kagan:2009bn}).  However,
significant differences from a linear expansion may arise if both top-
and bottom-quark Yukawa couplings are order one as it happens, for
instance, in the large $\tan\beta$ regime.

\subsection{MFV at large $\tan\beta$.}
\label{eq:largetanb}
If the Yukawa Lagrangian contains more than a single Higgs field, we
can still assume that the Yukawa couplings are the only irreducible
breaking sources of $SU(3)^3_{q}$, but we can change their overall
normalization.

A particularly interesting scenario is the two-Higgs-doublet model
where the two Higgs fields, $\phi_U$ and $\phi_D$, are coupled
separately to up- and down-type quarks:
\begin{equation}\label{eq:LY2}
- \mathcal{L}^{\rm 2HDM}_{\rm Yukawa}  = Y^d_{ij}~{\Qbar_{Li}}\phi_D
D_{Rj}
+Y^u_{ij}~{\Qbar_{Li}} \phi_U U_{Rj}+Y^e_{ij}~\Lbar_{Li}\phi_D E_{Rj}
+{\rm h.c.}.
\end{equation}
This Lagrangian is invariant under an extra ${\rm U}(1)$ symmetry with
respect to the one-Higgs Lagrangian in Eq.~(\ref{Hqint}): a symmetry
under which the only charged fields are $D_R$ and $E_R$ (charge $+1$)
and $\phi_D$ (charge $-1$).  This symmetry, denoted ${\rm U}(1)_{\rm
  PQ}$, prevents tree-level FCNCs and implies that $Y^{u,d}$ are the
only sources of $SU(3)^3_{q}$ breaking appearing in the Yukawa
interaction (similar to the one-Higgs-doublet scenario). Consistently
with the MFV hypothesis, we can then assume that $Y^{u,d}$ are the
only relevant sources of $SU(3)_{q}^3$ breaking appearing in all the
low-energy effective operators.  This is sufficient to ensure that
flavor-mixing is still governed by the CKM matrix, and naturally
guarantees a good agreement with present data in the $\Delta F =2$
sector. However, the extra symmetry of the Yukawa interaction allows
us to change the overall normalization of $Y^{u,d}$ with interesting
phenomenological consequences in specific rare modes.
These effects are related only to the large value of bottom Yukawa,
and indeed can be found also in other NP frameworks where there
is no extended Higgs sector, but the  bottom Yukawa coupling
is of order one~\cite{Kagan:2009bn}.

Assuming the Lagrangian in Eq.~(\ref{eq:LY2}), the normalization of
the Yukawa couplings is controlled by the ratio of the vacuum
expectation values of the two Higgs fields, or by the parameter
\be
\tan\beta = \langle \phi_U\rangle/\langle \phi_D\rangle~.
\ee
For $\tan\beta\gg1 $ the smallness of the $b$ quark and $\tau$ lepton
masses can be attributed to the smallness of $1/\tan\beta$ rather than
to the corresponding Yukawa couplings.  As a result, for
$\tan\beta\gg1$ we cannot anymore neglect the down-type Yukawa
coupling.  Moreover, the ${\rm U}(1)_{\rm PQ}$ symmetry cannot be
exact: it has to be broken at least in the scalar potential in order
to avoid the presence of a massless pseudoscalar Higgs.  Even if the
breaking of ${\rm U}(1)_{\rm PQ}$ and $SU(3)^3_{q}$ are decoupled, the
presence of ${\rm U}(1)_{\rm PQ}$ breaking sources can have important
implications on the structure of the Yukawa interaction, especially if
$\tan\beta$ is large~\cite{Hall:1993gn,Blazek:1995nv,
  Isidori:2001fv,D'Ambrosio:2002ex}.  We can indeed consider new
dimension-four operators such as
\begin{equation}
 \epsilon~ \Qbar_L  Y^d D_R  \tilde \phi_U
\qquad {\rm or} \qquad
 \epsilon~ \Qbar_L  Y^uY^{u\dagger} Y^d D_R  \tilde \phi_U~,
\label{eq:O_PCU}
\end{equation}
where $\epsilon$ denotes a generic MFV-invariant ${\rm U}(1)_{\rm
  PQ}$-breaking source. Even if $\epsilon \ll 1 $, the product
$\epsilon \times \tan\beta$ can be $\mathcal{O}(1)$, inducing large
corrections to the down-type Yukawa sector:
\begin{equation}
 \epsilon~ \Qbar_L  Y^d D_R  \tilde \phi_U
\ \stackrel{vev}{\longrightarrow}  \
 \epsilon~  \Qbar_L  Y^d D_R   \langle \tilde \phi_U \rangle =
  (\epsilon\times\tan\beta)~  \Qbar_L  Y^d D_R   \langle \phi_D \rangle~.
\end{equation}

Since the $b$-quark Yukawa coupling becomes $\mathcal{O}(1)$, the
large-$\tan\beta$ regime is particularly interesting for
helicity-suppressed observables in $B$ physics.  One of the clearest
phenomenological consequences is a suppression (typically in the
$10-50\%$ range) of the $B \to \ell \nu$ decay rate with respect to
its SM expectation~\cite{Hou:1992sy}.  Potentially measurable effects
in the $10-30\%$ range are expected also in $B\to X_s
\gamma$~\cite{Carena:1999py} and $\Delta M_{B_s}$~\cite{Buras:2001mb}.
Given the present measurements of $B \to \ell \nu$, $B\to X_s \gamma$,
and $\Delta M_{B_s}$, none of these effects seems to be favored by
data.  However, present errors are still sizable compared to the
estimated NP effects.

The most striking signature could arise from the rare decays
$B_{s,d}\to \ell^+\ell^-$, whose rates could be enhanced over the SM
expectations by more than one order of
magnitude~\cite{Hamzaoui:1998nu}.  An enhancement of both $B_{s}\to
\ell^+\ell^-$ and $B_{d}\to \ell^+\ell^-$ respecting the MFV relation
$\Gamma(B_{s}\to \ell^+\ell^-)/\Gamma(B_{d}\to \ell^+\ell^-) \approx
|V_{ts}/V_{td}|^2$ would be an unambiguous signature of MFV at large
$\tan\beta$~\cite{Hurth:2008jc}.

\subsection{MFV with additional flavor-diagonal phases}
\label{GMFV}
The breaking of the $SU(3)_q^3$ flavor group and the breaking of the
discrete CP symmetry are not necessarily related, and we can add
flavor-diagonal CPV phases to generic MFV
models~\cite{Ellis:2007kb,Mercolli:2009ns,Colangelo:2008qp}. Because of the
experimental constraints on electric dipole moments (EDMs), which are
generally sensitive to such flavour-diagonal
phases~\cite{Mercolli:2009ns}, in this more general case the bounds on
the scale of new physics are substantially higher with respect to the
``minimal'' case, where the Yukawa couplings are assumed to be the
only breaking sources of both symmetries~\cite{D'Ambrosio:2002ex}

If $\tan\beta$ is large, the inclusion of flavor-diagonal phases has
interesting effects also in flavour-changing processes. The main
consequences, derived in a model independent manner, can be summarized
as follows~\cite{Kagan:2009bn}: (i) extra CPV can only arise from
flavor diagonal CPV sources in the UV theory; (ii) the extra CP phases
in $B_s-\bar B_s$ mixing provide an upper bound on the amount of CPV
in $B_d-\bar B_d$ mixing; (iii) if operators containing right-handed
light quarks are sub-dominant then the extra CPV is equal in the two
systems, and is negligible in transitions between the first two
generations to the third one. Conversely, these operators can break
the correlation between CPV in the $B_s$ and $B_d$ systems, and can
induce significant new CPV in $\epsilon_K$.

\section{Supersymmetry}
\label{sec:susy}
Supersymmetric models provide, in general, new sources of flavor
violation, for both the quark and the lepton sectors. The main new
sources are the supersymmetry breaking soft mass terms for squarks and
sleptons, and the trilinear couplings of a Higgs field with a
squark-antisquark, or slepton-antislepton pairs. Let us focus on the
squark sector. The new sources of flavor violation are most commonly
analyzed in the basis in which the corresponding (down or up) quark
mass matrix and the neutral gaugino vertices are diagonal. In this
basis, the squark masses are not necessarily flavor-diagonal, and have
the form
\beq
\tilde q_{Mi}^*(M_{\tilde q}^2)^{MN}_{ij}\tilde q_{Nj}=
(\tilde q_{Li}^*\ \tilde q_{Rk}^*)\left(\begin{array}{cc}
    (M^2_{\tilde q})_{Lij} & A^q_{il}v_q \cr A^q_{jk}v_q &
    (M^2_{\tilde q})_{Rkl} \cr
  \end{array}\right) \left(\begin{array}{c}
    \tilde q_{Lj} \cr \tilde q_{Rl} \cr
  \end{array}\right),
\eeq
where $M,N=L,R$ label chirality, and $i,j,k,l=1,2,3$ are generation
indices. $(M^2_{\tilde q})_L$ and $(M^2_{\tilde q})_R$ are the
supersymmetry-breaking squark masses-squared. The $A^q$ parameters
enter in the trilinear scalar couplings $A^q_{ij}\phi_q\widetilde
q_{Li}\widetilde q_{Rj}^*$, where $\phi_q$ $(q=u,d)$ is the $q$-type
Higgs boson and $v_q=\langle\phi_q\rangle$.

In this basis, flavor violation takes place through one or more squark
mass insertion. Each mass insertion brings with it a factor of
$(\delta_{ij}^q)_{MN}\equiv(M^2_{\tilde q})^{MN}_{ij}/\tilde m_q^2$,
where $\tilde m^2_q$ is a representative $q$-squark mass
scale. Physical processes therefore constrain
\beq
[(\delta^q_{ij})_{MN}]_{\rm eff}\sim{\rm max}[(\delta^q_{ij})_{MN},
(\delta^q_{ik})_{MP}(\delta^q_{kj})_{PN},\ldots,(i\leftrightarrow j)].
\eeq
For example,
\beq
[(\delta^d_{12})_{LR}]_{\rm eff}\sim{\rm max}[A^d_{12}v_d/\tilde
m_d^2, (M^2_{\tilde d})_{L1k}A^d_{k2}v_d/\tilde
m_d^4,A^d_{1k}v_d(M^2_{\tilde d})_{Rk2}/\tilde
m_d^4,\ldots,(1\leftrightarrow2)].
\eeq
Note that the contributions with two or more insertions may be less
suppressed than those with only one.

In terms of mass basis parameters, the $(\delta^q_{ij})_{MM}$'s stand
for a combination of mass splittings and mixing angles:
\beq
(\delta^q_{ij})_{MM}=\frac{1}{\tilde m_{q}^2}\sum_\alpha
(K^q_M)_{i\alpha}(K^q_M)_{j\alpha}^*\Delta\tilde m^2_{q_\alpha},
\eeq
where $K^q_M$ is the mixing matrix in the coupling of the gluino (and
similarly for the bino and neutral wino) to $q_{Li}-\tilde
q_{M\alpha}$; $\tilde m^2_q=\frac13\sum_{\alpha=1}^3\tilde
m_{q_{M\alpha}}^2$ is the average squark mass-squared, and
$\Delta\tilde m^2_{q_\alpha}=\tilde m^2_{q_\alpha}-\tilde
m^2_q$. Things simplify considerably when the two following conditions
are satisfied \cite{Hiller:2008sv}, which means that a two generation
effective framework can be used (for simplicity, we omit here the
chirality index):
\beq
|K_{ik}K_{jk}^*|\ll|K_{ij}K_{jj}^*|,\ \ \
|K_{ik}K_{jk}^*\Delta\tilde
m^2_{q_jq_i}|\ll|K_{ij}K_{jj}^*\Delta\tilde m^2_{q_jq_i}|,
\eeq
where there is no summation over $i,j,k$ and where $\tilde
m^2_{q_jq_i}=\tilde m^2_{q_j}-\tilde m^2_{q_i}$. Then, the
contribution of the intermediate $\tilde q_k$ can be neglected and,
furthermore, to a good approximation,
$K_{ii}K_{ji}^*+K_{ij}K_{jj}^*=0$. For these cases, we obtain
\beq\label{eq:delmass}
(\delta^q_{ij})_{MM}=\frac{\Delta\tilde m^2_{q_jq_i}}{\tilde m_{q}^2}
(K^q_M)_{ij}(K^q_M)_{jj}^*.
\eeq
It is further useful to use instead of $\tilde m_q$ the mass scale
$\tilde m^q_{ij}=\frac12(\tilde m_{q_i}+\tilde m_{q_j})$ \cite{Raz:2002zx}.
We also define
\beq
\langle\delta^q_{ij}\rangle=\sqrt{(\delta_{ij}^{q})_{LL}
  (\delta^{q}_{ij})_{RR}}.
\eeq

The new sources of flavor and CP violation contribute to FCNC
processes via loop diagrams involving squarks and gluinos (or
electroweak gauginos, or higgsinos). If the scale of the soft
supersymmetry breaking is below TeV, and if the new flavor violation
is of order one, and/or if the phases are of order one, then these
contributions could be orders of magnitude above the experimental
bounds. Imposing that the supersymmetric contributions do not exceed
the phenomenological constraints leads to constraints of the form
$(\delta^q_{ij})_{MM}\ll1$. Such constraints imply that either
quasi-degeneracy ($\Delta\tilde m^2_{q_jq_i}\ll\tilde m^{q2}_{ij}$) or
alignment ($|K^q_{ij}|\ll1$) or a combination of the two mechanisms is
at work.

Table \ref{tab:exp} presents the constraints obtained in Refs.
\cite{Masiero:2005ua,Ciuchini:2007cw,Buchalla:2008jp,Gedalia:2009kh}
as appear in
\cite{Hiller:2008sv}. Wherever relevant, a phase suppression of order
0.3 in the mixing amplitude is allowed, namely we quote the stronger
between the bounds on ${\cal R}e(\delta^q_{ij})$ and $3{\cal
  I}m(\delta^q_{ij})$.  The dependence of these bounds on the average
squark mass $m_{\tilde q}$, the ratio $x\equiv m_{\tilde
  g}^2/m_{\tilde q}^2$ as well as the effect of arbitrary strong CP
violating phases can be found in \cite{Hiller:2008sv}.

\begin{table}[t]
\begin{center}
\begin{tabular}{cc|cc} \hline\hline
\rule{0pt}{1.2em}%
$q$\ & $ij\ $\ &  $(\delta^{q}_{ij})_{MM}$ &
$\langle\delta^q_{ij}\rangle$ \cr \hline
$d$ & $12$\ & $\ 0.03\ $ & $\ 0.002\ $ \cr
$d$ & $13$\ & $\ 0.2\ $ & $\ 0.07\ $ \cr
$d$ & $23$\ & $\ 0.6\ $ & $\ 0.2\ $ \cr
$u$ & $12$\ & $\ 0.1\ $ & $\ 0.008\ $ \cr
\hline\hline
\end{tabular}
\caption{The phenomenological upper bounds on $(\delta_{ij}^{q})_{MM}$ and
   on $\langle\delta^q_{ij}\rangle$, where $q=u,d$ and $M=L,R$.
   The constraints are given for $m_{\tilde q}=1$ TeV and $x\equiv m_{\tilde
   g}^2/m_{\tilde q}^2=1$. We assume that the phases could suppress the
   imaginary parts by a factor $\sim0.3$. The bound on
   $(\delta^{d}_{23})_{RR}$ is about 3 times weaker than that on
   $(\delta^{d}_{23})_{LL}$ (given in table). The constraints on
   $(\delta^{d}_{12,13})_{MM}$, $(\delta^{u}_{12})_{MM}$ and
   $(\delta^{d}_{23})_{MM}$ are based on, respectively,
   Refs. \cite{Masiero:2005ua}, \cite{Ciuchini:2007cw} and
   \cite{Buchalla:2008jp}. \label{tab:exp}}
\end{center}
\end{table}

For large $\tan\beta$, some constraints are modified from those in
Table~\ref{tab:exp}. For instance, the effects of neutral Higgs
exchange in $B_s$ and $B_d$ mixing give, for $\tan \beta =30$ and
$x=1$ (see \cite{Hiller:2008sv,Foster:2006ze} and references
therein for details):
\beq \label{eq:bmixbounds}
\langle \delta^d_{13}\rangle  < 0.01 \cdot \left( \frac{M_{A^0}}{200
    \, \mbox{GeV}} \right) , ~~~~~
\langle \delta^d_{23} \rangle < 0.04 \cdot \left( \frac{M_{A^0}}{200
    \, \mbox{GeV}} \right) ,
\eeq
where $M_{A^0}$ denotes the pseudoscalar Higgs mass, and the above
bounds scale roughly as $(30/\tan \beta)^2$.

The experimental constraints on the $(\delta^q_{ij})_{LR}$ parameters
in the quark-squark sector are presented in Table~\ref{tab:expLRme}.
The bounds are the same for $(\delta^q_{ij})_{LR}$ and
$(\delta^q_{ij})_{RL}$, except for $(\delta^d_{12})_{MN}$, where the
bound for $MN=LR$ is 10 times weaker.  Very strong constraints apply
for the phase of $(\delta^q_{11})_{LR}$ from EDMs. For $x=4$ and a
phase smaller than 0.1, the EDM constraints on
$(\delta^{u,d,\ell}_{11})_{LR}$ are weakened by a factor $\sim6$.

\begin{table}[t]
\label{tab:expLRme}
\begin{center}
\begin{tabular}{cc|c} \hline\hline
\rule{0pt}{1.2em}%
$q$\ & $ij$\ & $(\delta^{q}_{ij})_{LR}$\cr \hline
$d$ & $12$\  &    $\ 2\times10^{-4} \ $ \cr
$d$ & $13$\  &  $\ 0.08 \ $  \cr
$d$ & $23$\  &  $\ 0.01 \ $ \cr
$d$ & $11$\  & $4.7\times10^{-6}$ \cr
$u$ & $11$\  &  $9.3\times 10^{-6}$ \cr
$u$ & $12$\  & $\ 0.02 \ $\cr
\hline\hline
\end{tabular}
\caption{The phenomenological upper bounds on chirality-mixing
  $(\delta_{ij}^{q})_{LR}$, where $q=u,d$. The constraints are
   given for $m_{\tilde q}=1$ TeV and $x\equiv m_{\tilde
   g}^2/m_{\tilde q}^2=1$.  The constraints on
   $\delta^{d}_{12,13}$, $\delta^{u}_{12}$, $\delta^{d}_{23}$ and
   $\delta^{q}_{ii}$ are based on, respectively,
   Refs.~\cite{Masiero:2005ua}, \cite{Ciuchini:2007cw},
   \cite{Buchalla:2008jp} and \cite{Raidal:2008jk} (with the relation
   between the neutron and quark EDMs as in \cite{Gabbiani:1996hi}).}
\end{center}
\end{table}

While, in general, the low energy flavor measurements constrain only
the combinations of the suppression factors from degeneracy and from
alignment, such as Eq. (\ref{eq:delmass}), an interesting exception
occurs when combining the measurements of $K^0$--$\Knotbar$ and
$D^0$--$\Dnotbar$ mixing to test the first two generation squark
doublets. Here, for masses below the TeV scale, some level of
degeneracy is unavoidable \cite{Blum:2009sk}:
\beq
\frac{m_{\widetilde Q_2}-m_{\widetilde Q_1}}{m_{\widetilde
    Q_2}+m_{\widetilde Q_1}}\leq\begin{cases} 0.034 & {\rm maximal\
    phases} \cr  0.27 & {\rm vanishing\ phases}\cr \end{cases}
\eeq

The strong constraints in Tables \ref{tab:exp} and \ref{tab:expLRme}
can be satisfied if the mediation of supersymmetry breaking to the
MSSM is MFV. In particular, if at the scale of mediation, the
supersymmetry breaking squark masses are universal, and the A-terms
vanish or are proportional to the Yukawa couplings, then the model is
phenomenologically safe. Indeed, there are several known mechanisms of
mediation that are MFV (see, {\it e.g.} \cite{Shadmi:1999jy}). In
particular, gauge-mediation
\cite{Dine:1993yw,Dine:1994vc,Dine:1995ag,Meade:2008wd},
anomaly-mediation \cite{Randall:1998uk,Giudice:1998xp}, and
gaugino-mediation \cite{Chacko:1999mi} are such mechanisms. (The
renormalization group flow in the MSSM with generic MFV soft-breaking
terms at some high scale has recently been discussed in
Ref.~\cite{Colangelo:2008qp,Paradisi:2008qh}.) On the other hand, we
do not expect gravity-mediation to be MFV and it could provide
subdominant, yet observable flavor and CP violating effects
\cite{Feng:2007ke}.

\section{Extra Dimensions}
\label{sec:exdim}
Models of extra dimensions come in a large variety and the
corresponding phenomenology, including the implications for flavor
physics, changes from one extra dimension framework to another.  Yet,
as in the supersymmetric case, one can classify the new sources of
flavor violation which generically arise:

{\bf Bulk masses} - If the SM fields propagate in the bulk of the
extra dimensions they can have bulk, vector-like, masses. These mass
terms are of particular importance to flavor physics since they induce
fermion localization which may yield hierarchies in the low energy
effective couplings.  Furthermore, the bulk masses, which define the
extra dimension interaction basis, do not need to commute with the
Yukawa matrices, and hence might induce contributions to FCNC
processes, similarly to the squark soft masses-squared in
supersymmetry.

{\bf Cutoff, UV physics} - Since, generically, higher dimensional
field theories are non-renormalizable, they rely on unspecified
microscopic dynamics to provide UV completion of the models. Hence,
they can be viewed as effective field theories and the impact of the
UV physics is expected to be captured by a set of operators suppressed
by the, framework dependent, cutoff scale.  Without precise knowledge
of the short distance dynamics, the additional operators are expected
to carry generic flavor structure and contribute to FCNC processes.
This is somewhat similar to ``gravity mediated'' contributions to
supersymmetry-breaking soft terms which are generically expected to
have an anarchic flavor structure and are suppressed by the Planck scale.

{\bf ``Brane'' localized terms} - The extra dimensions have to be
compact and typically contain defects and boundaries of smaller
dimensions [in order, for example, to yield a chiral low energy four
dimension (4D) theory].  These special points might contain different
microscopical degrees of freedom. Therefore, generically, one expects
that a different and independent class of higher dimension operators
may be localized to this singular region in the extra dimension
manifold. (These are commonly denoted `brane terms' even though, in
most cases, they have very little to do with string theory). The
brane-localized terms can, in principle, be of anarchic flavor
structure and provide new flavor and CP violating sources.  One
important class of such operators are brane kinetic terms: their
impact is somewhat similar to that of non-canonical kinetic terms
which generically arise in supersymmetric flavor models.

We focus on flavor physics of five dimension (5D) models, with bulk SM
fields, since most of the literature focuses on this class.
Furthermore, the new flavor structure that arises in 5D models
captures most of the known effects of extra-dimension flavor models.
Assuming a flat extra dimension, the energy range, $\Lambda_{\rm 5D}
R$ (where $\Lambda_{\rm 5D}$ is the 5D effective cutoff scale and $R$
is the extra dimension radius with the extra dimension coordinate
$y\in(0,\pi R)$), for which the 5D effective field theory holds, can be
estimated as follows. Since gauge couplings in extra dimensional
theories are dimensional, {\it i.e.}~$\alpha_{\rm 5D}$ has mass
dimension $-1$, a rough guess (which is confirmed, up to order one
corrections, by various NDA methods) is~\cite{Kribs:2006mq}
$\Lambda_{\rm 5D} \sim 4 \pi/{\alpha_{\rm 5D}} \,.$ Matching this 5D
gauge coupling to a 4D coupling of the SM, at leading order,
${1}/{g^2} = \pi R/{g_{\rm 5D}^2}\,,$ we obtain
\begin{equation}
\Lambda_{\rm 5D} R \sim \frac{4}{\alpha} \sim 30\,.
\end{equation}
Generically, the mass of the lightest Kaluza-Klein (KK) states,
$\Mkk$, is of ${\cal O}\big(R^{-1}\big)$.  If the extra dimension
theory is linked to the solution of the hierarchy problem and/or
directly accessible to near future experiments, then $R^{-1}= {\cal
  O}\big(\rm TeV\big)$.  This implies an upper bound on the 5D cutoff:
\begin{equation}
\Lambda_{\rm 5D}\lesssim 10^2\,{\rm TeV}\ll \Lambda_K\sim 2 \times
10^5\,\rm TeV \,,
\end{equation}
where $ \Lambda_K$ is the scale required to suppress the generic
contributions to $\epsilon_K$, discussed above (see
Table~\ref{tab:DF2}).

The above discussion ignores the possibility of splitting the fermions
in the extra dimension. In split fermion models different bulk masses
are assigned to different generations. Consequently fermions are
localized and separated in the bulk of the extra dimension in a manner
which may successfully address the SM flavor
puzzle~\cite{ArkaniHamed:1999dc}.  Separation in the extra dimension
may suppress the contributions to $\epsilon_K$ from the
higher-dimension cut-off induced operators.  As shown in
Table~\ref{tab:DF2}, the most dangerous operator is
\be
{O^4_K} = \frac{1}{\Lambda_{\rm 5D}^2}  \left(\bar s_L \,d_R\right)
\left(\bar s_R\, d_L\right)~.
\label{eq:O4K}
\ee
This operator contains $s$ and $d$ fields of both chiralities.  As a
result, in a large class of split fermion models, the overlap
suppression would be similar to that accounting for the smallness of
the down and strange 4D Yukawa couplings.  Integrating over the 5D
profiles of the four quarks, this may yield a suppression factor of
${\cal O}\big(m_d m_s/v^2\big)\sim10^{-9}$. Together with the naive
scale suppression, $1/\Lambda_{\rm 5D}^2$, the coefficient of
${O^4_K}$ can be sufficiently suppressed to be consistent with the
experimental bound.

In the absence of large brane kinetic terms (BKTs), however, fermion
localization generates order one non-universal couplings to the gauge
KK fields~\cite{Delgado:1999sv}. (See {\it e.g.}~\cite{AgasheTasi} and
references therein. The case with large BKTs is similar to the warped
case discussed below.) The fact that the bulk masses are, generically,
not aligned with the 5D Yukawa couplings implies that KK gluon
exchange processes induce, among others, the following operator in the
low energy theory: $\left[(D_L)_{12}^2/(6\Mkks)\right]\left(\bar s_L\,
  d_L\right)^2$, where $\left(D_{L}\right)_{12}\sim \lambda$ is the
left-handed down-quark rotation matrix from the 5D interaction basis
to the mass basis. This structure provides only a mild suppression to
the resulting operator. It implies that to satisfy the $\epsilon_K$
constraint the KK and the inverse compactification scales have to be
above $10^3$\,TeV, beyond the direct reach of near future experiments,
and too high to be linked to a solution of the hierarchy problem.
This problem can be solved by tuning the 5D flavor parameters and
imposing appropriate 5D flavor symmetries to make the tuning stable.
Once the 5D bulk masses are aligned with the 5D Yukawa matrices the KK
gauge contributions would vanish and such a configuration is
radiatively stable.

The warped extra-dimension [Randall-Sundrum (RS)]
framework~\cite{Randall:1999ee} provides a solution to the hierarchy
problem. Moreover, with SM fermions propagating in the bulk, both the
SM and the NP flavor puzzles can be addressed.  The light fermions
could be localized away from the TeV brane~\cite{GN}, where the Higgs
is localized. Such a configuration can generate the observed Yukawa
hierarchy and, at the same time, ensure that higher-dimensional
operators are suppressed by a high cutoff scale associated with the
location of the light fermions in the extra
dimension~\cite{RSoriginal}.  Furthermore, since the KK states are
localized near the TeV brane, the couplings between the SM quarks and
the gauge KK fields exhibit the hierarchical structure associated with
SM masses and CKM mixings.  This hierarchy in the couplings provides
an extra protection against non-standard flavor-violating
effects~\cite{Huber:2003tu} denoted as RS-GIM mechanism~\cite{aps}
(see also~\cite{Burdman:2002gr}). It is interesting to note that an
analogous mechanism is at work in models with strong dynamics at the
TeV scale, with large anomalous dimension and partial
compositeness~\cite{GK}. The link with strongly-interacting models in
indeed motivated by the AdS/CFT correspondence~\cite{AdSCFT}, which
implies that the above 5D framework is the dual description of 4D
composite Higgs models~\cite{AdSCFTpheno}.

Concerning the quark zero modes, the flavor structure of the above
models as well as the phenomenology can be captured by using the
following simple rules~\cite{aps,Gedalia:2009ws,Contino:2006nn}. In
the 5D interaction basis, where the bulk masses $k\, C^{ij}_{x}$ are
diagonal ($x=Q,U,D$; $i,j=1,2,3$; $k$ is the AdS curvature), the
value $f_{x^i}$ of the profile of the quark zero modes is given by
\be \label{fs}
f_{x^i}^2=(1-2c_{x^i})/( 1-\epsilon^{ 1-2c_{x^i} })\,.
 \ee
 Here $c_{x^i}$ are the eigenvalues of the $C_x$ matrices,
 $\epsilon=\exp[-\xi]$, $\xi=\log[M_{\rm \overline{Pl}}/{\rm TeV}]$,
 and $M_{\rm \overline{Pl}}$ is the reduced Planck mass.  If
 $c_{x^i}<1/2$, then $f_{x^i}$ is exponentially suppressed. Hence,
 order one variations in the 5D masses yield large hierarchies in the
 4D flavor parameters.  We consider the cases where the Higgs VEV
 either propagates in the bulk~\cite{Agashe:2008uz} or is localized on
 the IR brane.  For a bulk Higgs case, the profile is given by
 $\tilde{v}( \beta, z) \simeq v\sqrt{k(1+\beta)} \bar z^{2+\beta}/
 \epsilon $, where $\bar z\in(\epsilon,1)$ ($\bar z=1$ on the IR brane),
 and $\beta\geq0$. The $\beta=0$ case describes a Higgs
 maximally-spread into the bulk (saturating the AdS stability
 bound~\cite{Breitenlohner:1982bm}).  The relevant part of the
 effective 4D Lagrangian, which involves the zero modes and the first
 KK gauge states can be approximated by~\cite{aps,Gedalia:2009ws}
\be \label{lagrangian}\hspace*{-.15cm}
\mathcal{L}^{4D} \supset  (Y^{u,d}_{\rm 5D})_{ij} \phi^{u,d} \,
\bar Q_i f_{Q_i} \left(U,D\right)_j f_{U_j,D_j} r^\phi_{00}
(\beta,c_{Q_i},c_{U_j,D_j}) +g_*  G^1 x^\dagger_i x_i
\left[f_{x^i}^2 \rg(c_{x^i}) -{1}/{\xi}  \right]  ,
\ee
where $\phi^{u,d}=\tilde \phi, \phi$, $g_* $ stands for a generic
effective gauge coupling and summation over $i,j$ is implied.  The
correction for the couplings from the case of fully IR-localized KK
and Higgs states is given by the functions
$r^\phi_{00}$~\cite{Gedalia:2009ws} and
$\rg$~\cite{cfw1,Csaki:2009bb}:
\be \label{r00}
r^\phi_{00}(\beta,c_L,c_R)
\approx \frac{\sqrt{2(1+\beta)}}{2+\beta-c_L-c_R} \,, \ \  \ \rg(c)
\approx \frac{\sqrt{2}}{J_1(x_1)} \frac{0.7}{6-4c} \left(
1+e^{c/2} \right) \,,
\ee
where $r^\phi_{00}(\beta,c_L,c_R)=1$ for brane-localized Higgs and
$x_1 \approx 2.4$ is the first root of the Bessel function,
$J_0(x_1)=0$.

In Table~\ref{fstab} we present an example of a set of
$f_{x^i}$-values that, starting from anarchical 5D Yukawa couplings,
reproduce the correct hierarchy of the flavor parameters.  We assume,
for simplicity, an IR localized Higgs. The values depend on two input
parameters: $f_{U^3}$, which has been determined assuming a maximally
localized $t_R$ ($c_{u^3}=-0.5$), and $\yfd$, the overall scale of the
5D Yukawa couplings in units of $k$, which has been fixed to its
maximal value assuming three KK states.  On general grounds, the value
of $\yfd$ is bounded from above, as a function of the number of KK
levels, by the requirement that Yukawa interactions are perturbative
below the cutoff of the theory, $\Lambda_{\rm 5D}\sim \Nkk k$, and it
is bounded from below in order to account for the large top mass.
Hence the following range for $\yfd$ is obtained (see {\it
  e.g.}~\cite{Agashe:2008uz,LCP}):
\begin{equation}
\frac{1}{ 2}\lesssim \yfd \lesssim \frac{2\pi}{ \Nkk} {\rm \  \ for \
  brane \ Higgs\,;} \ \ \ \ \
\frac{1}{ 2}\lesssim \yfd \lesssim \frac{4\pi}{ \sqrt\Nkk}  {\rm \  \
  for \ bulk \ Higgs\,,}
\label{lam5D}\end{equation}
where we use the rescaling $y_{\rm 5D}\to y_{\rm 5D}\,
\sqrt{1+\beta}$, which produces the correct $\beta\to \infty$
limit~\cite{HFCNC} and avoids subtleties in the $\beta=0$ case.

{\small
\begin{table}[t]\begin{center}
 \begin{tabular}{c|c|c|c}
    \hline\hline
    { Flavor}& { $f_Q$} & { $f_U$} & { $f_D$}\cr
    \hline
    1 &$ {{A \lambda^{3}} { f_{Q^3}}}\sim 3 \times
    10^{-3}$&
    $\frac{m_u}{m_t}\, \frac{ f_{U^3} }{ A \lambda^3} \sim1\times10^{-3}$&
     $\frac{m_d}{m_b}\, \frac{ f_{D^3} }{ A \lambda^3} \sim 2\times 10^{-3}$
    \cr
    2&$ {{ A \lambda^{2}} { f_{Q^3}}}\sim
    1\times 10^{-2}$&
    $\frac{ m_c }{ m_t} \, \frac{ f_{U^3} }{  A \lambda^2} \sim 0.1$&
      $\frac{ m_s}{ m_b}\, \frac{ f_{D^3} }{ A \lambda^2} \sim 1\times 10^{-2}$
    \cr
    3 &$ \frac{ m_t}{ v \yfd f_{U^3}}\sim 0.3$ &$\sqrt2$&
    $\frac{ m_b }{ m_t} \, { f_{U^3}}\sim 2\times 10^{-2}$
    \vspace*{.05cm}\cr
\hline\hline\end{tabular}
\caption{{\small Values of the $f_{x^i}$ parameters [Eq.~(\ref{fs})]
  which reproduce the observed quark masses and CKM mixing angles starting from
 anarchical 5D Yukawa couplings. We fix $f_{U^3}=\sqrt2$ and $\yfd=2$
 (see text).}}\label{fstab}
  \end{center}
\end{table}
}

\begin{table}[t]
\begin{center}
\begin{tabular}{c|c c|c c} \hline\hline
\rule{0pt}{1.2em}%
Observable &  \multicolumn{2}{c}{$M_G^{\rm min}[\mathrm{TeV}]$} &
\multicolumn{2}{|c}{$\yfdmin\ {\rm or}\ f_{Q_3}^{\rm max}$} \cr
&   IR Higgs& $\beta=0$ & IR Higgs & $\beta=0$ \cr
 \hline CPV-$B_d^{LLLL}$ &  $12 f_{Q_3}^2$ &$12 f_{Q_3}^2$&
 $f_{Q_3}^{\rm max}=0.5$ &$f_{Q_3}^{\rm max}=0.5$\cr
CPV-$B_d^{LLRR}$  & ${4.2/ \yfd}$ &${2.4/ \yfd}$&  $\yfdmin=1.4$ &$\yfdmin=0.82$ \cr
CPV-$D^{LLLL}$ &  $0.73 f_{Q_3}^2$ &$0.73 f_{Q_3}^2$& no bound & no bound\cr
CPV-$D^{LLRR}$   & ${4.9/ \yfd}$ &${2.4/ \yfd}$& $\yfdmin=1.6$ &$\yfdmin=0.8$ \cr
$\epsilon_K^{LLLL}$ &  $7.9 f_{Q_3}^2$& $7.9 f_{Q_3}^2$ &
$f_{Q_3}^{\rm max}=0.62$ &$f_{Q_3}^{\rm max}=0.62$\cr
$\epsilon_K^{LLRR}$  & ${49/ \yfd}$ &$ {24/ \yfd}$& above
(\ref{lam5D}) & $\yfdmin=8 $ \cr
 \hline
 \hline
\end{tabular}
\caption{Most significant flavor constraints in the RS framework.
The values of $\yfdmin$ and $ f_{Q_3}^{\rm max}$ correspond to $\Mkk=3$ TeV.
The bounds are obtained assuming maximal CPV phases and
$g_{s*}=3$. Entries marked `above (\ref{lam5D})' imply that for
$\Mkk=3$ TeV, $\yfd$ is outside the perturbative range. \label{tab:rszi}}
\end{center}
\end{table}

With anarchical 5D Yukawa matrices, an RS residual little CP problem
remains~\cite{LCP}: Too large contributions to the neutron electric
dipole moment (EDM)~\cite{aps}, and sizable chirally enhanced
contributions to
$\epsilon_K$~\cite{cfw1,Blanke:2008zb,Davidson:2007si,Bona:2007vi,RSC}
are predicted. The RS leading contribution to $\epsilon_K$ is
generated by a tree-level KK-gluon exchange which leads to an
effective coupling for the chirality-flipping operator in
(\ref{eq:O4K}) of the
type~\cite{cfw1,Blanke:2008zb,Davidson:2007si,RSC}
\bea\label{eq:C4K}
C_4^K &\simeq&\frac{g_{s*}^2}{M_{KK}^2} f_{Q_2} f_{Q_1} f_{d_2}
f_{d_1} \rg(c_{Q_2}) \rg(c_{d_2}) \no\\
&\sim& \frac{g_{s*}^2}{M_{KK}^2} \frac{2 m_d m_s }{(v \yfd)^2}
\frac{\rg(c_{Q_2}) \rg(c_{d_2})}{r^H_{00}(\beta,c_{Q_1},c_{d_1})
  r^H_{00}(\beta,c_{Q_2},c_{d_2})}\,.
\eea
The final expression is independent of the $f_{x^i}$, so the bound in
Table~\ref{tab:DF2} can be translated into constraints in the
$\yfd-\Mkk$ plane. The analogous effects in the $D$ and $B$ systems
yield numerically weaker bounds.  Another class of contributions,
which involves only left-handed quarks, is also important to constrain
the $f_Q-\Mkk$ parameter space.

In Table~\ref{tab:rszi} we summarize the resulting constraints. For
the purpose of a quantitative analysis we set $g_{s*}=3$, as obtained
by matching to the 4D coupling at one-loop~\cite{Agashe:2008uz} (for
the impact of a smaller RS volume see~\cite{Davoudiasl:2008hx}).  The
constraints related to CPV correspond to maximal phases, and are
subject to the requirement that the RS contributions are smaller than
$30\%$ ($60\%$) of the SM contributions~\cite{NMFV} in the $B_d$ ($K$)
system.  The analytical expressions in the table have roughly a 10\%
accuracy over the relevant range of parameters.  Contributions from
scalar exchange, either Higgs~\cite{HFCNC} or
radion~\cite{Azatov:2008vm}, are not included since these are more
model dependent and known to be weaker~\cite{Duling:2009pj} in the brane localized Higgs case.

Constraints from $\epsilon'/\epsilon_K$ have a different parameter
dependence than the $\epsilon_K$ constraints. Explicitly, for
$\beta=0$, the $\epsilon'/\epsilon_K$ constraint reads $M_G^{\rm
  min}=1.2\yfd$ TeV. When combined with the $\epsilon$ constraint, we
find $M_G^{\rm min}=5.5$ TeV with a corresponding
$\yfdmin=4.5$~\cite{Gedalia:2009ws}.

The constraints summarized in Table~\ref{tab:rszi} and the
contributions to the neutron EDM which generically require $\Mkk>
{\cal O}\left(20\,\rm TeV\right)$~\cite{aps} are a clear manifestation
of the RS little CP problem.  The problem can be amended by various
alignment mechanisms~\cite{LCP,Csaki:2009bb,Align}.  In this case the
bounds from the up sector, especially from CPV in the $D$
system~\cite{Blum:2009sk,Gedalia:2009kh}, become important.
Constraint from $\Delta F=1$ processes (in either the down
sector~\cite{aps,RSC} or $t\to c Z$~\cite{Agashe:2006wa}) are not
included here, since they are weaker and, furthermore, these
contributions can be suppressed (see~\cite{RSC}) due to incorporation of a
custodial symmetry~\cite{Agashe:2006at}.

\section{Future Prospects}
\label{sec:future}
The new physics flavor puzzle is the question of why, and in what way,
the flavor structure of TeV-scale new physics is non-generic. Indeed,
the flavor predictions of most new physics models are not a
consequence of their generic features but rather of the special
structures that are imposed specifically to satisfy the existing
severe flavor bounds.  Therefore, flavor physics is a powerful
indirect probe of new physics.  We hope that new physics not far above
the weak scale will be discovered at the LHC. A major issue will then
be to understand its flavor structure. While it is not easy to
directly probe this flavor structure at high energy, a lot can be
learned from low energy flavor physics.

The precision with which we can probe high scale physics in flavor
physics experiments is often limited by theoretical uncertainties.
Moreover, in case of theoretically clean observables, the sensitivity
to the new-physics scale increases slowly with the statistics of
the experiment. Thus, the important questions in view of
future experiments are the following:
\begin{enumerate}
\item
What are the expected deviations from the SM predictions induced by
new physics at the TeV scale?
\item
Which observables are not limited by theoretical uncertainties?
\item
In which case we can expect a substantial improvement on the
experimental side?
\item
What will the measurements teach us if deviations from the SM are
[not] seen?
\end{enumerate}
These questions have been analysed in a series of recent works (see
e.g.~\cite{Raidal:2008jk,Buchalla:2008jp,Grossman:2009dw,Antonelli:2009ws,Buras:2009if})
and the main conclusions can be summarized as follows:
\begin{enumerate}
\item
The expected deviations from the SM predictions induced by new physics at the
TeV scale with generic flavor structure are already ruled out by many
orders of magnitudes. On general grounds, we can expect any size of deviation
below the current bounds. In the most pessimistic frameworks, such as MFV,
the typical size of the deviations is at the few \% level
in FCNC amplitudes.
\item
The theoretical limitations are highly process dependent.
In most multi-hadron final states the calculation of
decay amplitudes are already limited by theoretical
uncertainties. However,
several channels involving leptons in the final state,
and selected time-dependent asymmetries, have a theoretical
errors well below the current experimental sensitivity.
\item
On the experimental side there are good prospect of improvements.
As summarized in Table~\ref{tab:future}, one order of magnitude
improvements in several clean $B_{s,d}$, $D$, and $K$ observables
are possible within a few years. Moreover, improvements of
several orders of magnitudes are expected in
top decays, which will be explored for the first
time in great detail at the LHC.
\item
There is no doubt that new low-energy flavor data will be
complementary with the high-$p_T$ part of the LHC program.
As illustrated in the previous sections,
the synergy of both data sets can teach us a lot
about the new physics at the TeV scale.
\end{enumerate}

{\small
\begin{table}[pt]
\begin{tabular}{l|c|c|c|c|l}\hline
  & SM
  & Theory   &   Present  & Future  &  Future  \\ [-8 pt]
\raisebox{10pt}{Observable}  &  prediction
 & error &  result     &  error  & Facility
\\ \hline\hline
$|V_{us}|\quad$ [$K\to\pi\ell\nu$] &  input  &  $0.5\% \to 0.1\%_{\rm Latt}$   &  $0.2246\pm0.0012$
  &  0.1\%   & {\footnotesize $K$ factory}  \\
$|V_{cb}|\quad$ [$B\to X_c \ell \nu$]  &  input  & $1\%$ &  $(41.54 \pm 0.73) \times 10^{-3}$
  &  $1\%$ & {\footnotesize Super-$B$}   \\
$|V_{ub}|\quad$ [$B\to \pi\ell\nu$]  &  input  &  $10\% \to 5\%_{\rm Latt}$  &  $(3.38 \pm 0.36)\times 10^{-3}$
  &  4\%   & {\footnotesize Super-$B$ }  \\
$\gamma\qquad\ $  [$B \to DK$]  &  input  & $<1^\circ$ &  $(70^{+27}_{-30})^\circ$
  & $3^\circ$ & {\footnotesize LHCb} \\ \hline
$S_{B_d \to \psi K}$  &  $\sin(2\beta)$ &  $\lsim 0.01$  &  $0.671 \pm 0.023$
  &  0.01 &  {\footnotesize LHCb}   \\
$S_{B_s\to \psi\phi}$  &   0.036   &  $\lsim 0.01$ &
  $0.81^{+0.12}_{-0.32}$  &  $0.01$  &  {\footnotesize LHCb} \\
$S_{B_d \to \phi K}$  &  $\sin(2\beta)$   &  $\lsim 0.05$ &  $0.44 \pm 0.18$
  &  0.1  & {\footnotesize LHCb} \\
$S_{B_s\to \phi\phi}$  &  $0.036$ &    $\lsim 0.05$ & ---
  &  $0.05$ & {\footnotesize LHCb} \\
$S_{B_d \to K^* \gamma}$  &  few $\times$ 0.01  &  $0.01$  & $ -0.16 \pm 0.22$
  &  $0.03$  & {\footnotesize Super-$B$}    \\
$S_{B_s\to\phi\gamma}$  &  few $\times$ 0.01  &   $0.01$ & ---
  &  $0.05$ & {\footnotesize LHCb}   \\
$A_{\rm SL}^d$  &  $-5 \times 10^{-4}$  &   $10^{-4}$  & $-(5.8 \pm 3.4) \times 10^{-3}$
  &   $10^{-3}$ & {\footnotesize LHCb}   \\
$A_{\rm SL}^s$  &  $2 \times 10^{-5}$  &   $< 10^{-5}$   & $(1.6 \pm 8.5) \times 10^{-3}$
  &   $10^{-3}$ & {\footnotesize LHCb}   \\ \hline
$A_{CP}(b\to s\gamma)$  &  $<0.01$  &   $<0.01$ & $-0.012\pm0.028$  &  0.005  & {\footnotesize Super-$B$} \\
$\cB(B\to \tau\nu)$  &  $1\times 10^{-4}$  &  $20\% \to 5\%_{\rm Latt}$ & $(1.73\pm0.35)\times10^{-4}$
  &  5\%  & {\footnotesize Super-$B$}    \\
$\cB(B\to \mu\nu)$  &  $4\times 10^{-7}$  &  $20\% \to 5\%_{\rm Latt}$ & $<1.3\times10^{-6}$
  &  6\%   & {\footnotesize Super-$B$}   \\
$\cB(B_s\to \mu^+\mu^-)$  &  $3\times 10^{-9}$  &  $20\% \to 5\%_{\rm Latt}$ & $<5\times10^{-8}$
  &  $10\%$  & {\footnotesize LHCb} \\
$\cB(B_d\to \mu^+\mu^-)$  &  $1\times 10^{-10}$  & $20\% \to 5\%_{\rm Latt}$ & $<1.5\times10^{-8}$
  &  [?]  & {\footnotesize LHCb} \\
$A_{\rm FB}(B\to K^*\mu^+\mu^-)_{q^2_0}$  &  0  &  $0.05$ & $(0.2 \pm 0.2)$
  &  $0.05$   & {\footnotesize LHCb}  \\
$B\to K\nu\bar\nu$  &  $4\times 10^{-6}$  &  $20\% \to 10\%_{\rm Latt}$ & $< 1.4\times 10^{-5}$
  &  20\%  & {\footnotesize Super-$B$} \\ \hline
$|q/p|_{D-{\rm mixing}}$ &  1  &   $< 10^{-3}$   & $(0.86^{+0.18}_{-0.15})$
  &   $0.03$ & {\footnotesize Super-$B$}   \\
$\phi_D\quad$  &  0  &   $< 10^{-3}$   & $−(9.6^{+8.3}_{-9.5})^\circ $
  &   $2^\circ$ & {\footnotesize Super-$B$}   \\ \hline
$\cB(K^+\to \pi^+\nu\bar\nu)$  &   $8.5 \times 10^{-11}$   & 8\% & $(1.73^{+1.15}_{-1.05})\times 10^{-10}$
 & 10\%  & {\footnotesize $K$ factory}   \\
$\cB(K_L \to \pi^0 \nu\bar\nu)$  &  $2.6 \times 10^{-11}$ & 10\% & $<2.6 \times 10^{-8}$  &  [?] & {\footnotesize $K$ factory}  \\
$ R^{(e/\mu)}(K \to \pi \ell \nu)$
& $2.477 \times 10^{-5}$  & 0.04\% &  $(2.498 \pm 0.014)\times 10^{-5}$   &  0.1\%  & {\footnotesize $K$ factory}   \\
\hline $\cB(t \to c\,Z, \gamma)$  &  ${\cal O}\left(10^{-13}\right)$ &   ${\cal O}\left(10^{-13}\right)$  &  $< 0.6 \times 10^{-2}$  & ${\cal O}\left(10^{-5}\right)$  & {\footnotesize LHC ($100$\,fb$^{-1}$)}  \\
\hline\hline
\end{tabular}\vspace*{4pt}
\caption{\small Status and prospects of selected $B_{s,d}$, $D$, $K$ and $t$ observables
(based on  information from Ref.~\cite{Buchalla:2008jp,Grossman:2009dw,Antonelli:2009ws}).
In the third column ``Latt'' refer to improvements in Lattice QCD expected in the next 5 years.
In the fourth column the bounds are 90\%\,CL. The errors in the fifth column
refer to  10\,fb$^{-1}$  at LHCb,  50\,ab$^{-1}$ at Super-$B$,
and two years at NA62 (``$K$ factory''). In the  third and  fifth column
the errors followed by ``\%" are relative errors, whil the others are absolute errors.
For entries marked ``[?]'' we have not found a reliable estimate of the
future experimental prospects. }
\label{tab:future}
\end{table}
}

While some improvements can still be expected from running
experiments, in particular from CDF and D0 at Fermilab, a substantial
step forward can be achieved only with the new dedicated facilities.
At the LHC, which has just started to operate, the LHCb experiment is
expected to collect about 10\,fb$^{-1}$ of data by 2015.  Beyond that,
an LHCb upgrade is planned with 10 times larger luminosity.  At CERN
the NA62 experiment is expected to start in 2012, with the main goal
of collecting $\cO(100)$ events of the rare decay $K^+ \to \pi^+
\nu\bar\nu$ by 2014. Dedicated rare $K$ experiments are planned also
at JParc and at Fermilab. The project to upgrade KEK-B into a
Super-$B$ factory, with a luminosity exceeding $5 \times 10^{35}/{\rm
  cm}^2s^{-1}$ has just started, and an even more ambitious super-$B$
project is currently under study in Italy.

In Table~\ref{tab:future}, based on data from
Ref.~\cite{Buchalla:2008jp,Grossman:2009dw,Antonelli:2009ws,TevT,Carvalho:2007yi}, we show
the possible improvements for a series of particularly significant
$B_{s,d}$, $D$, $K$ and $t$ observables, most of which have already been
discussed in the previous sections.  The future improvements refer to
10\,fb$^{-1}$ at LHCb, 50\,ab$^{-1}$ at Super-$B$, and two years of
nominal data taking at NA62. The table is not comprehensive and the
entries in the last two columns necessarily have significant
uncertainties. However, it illustrates two important points: i) there
are still several clean observables where we can expect significant
improvements in the near future; ii) progress in this field requires
the combined efforts of different experimental facilities.

\medskip

To conclude, we stress that the future of flavor physics is promising:
\begin{itemize}
\item The technology to collect much more flavor data exists.
\item There are still several measurements which are not theory limited.
\item Most well-motivated TeV-scale extensions of the Standard Model
  predict deviations within future experimental sensitivities and
  above the SM theoretical uncertainties.
\end{itemize}
The combination of direct discoveries at the LHC and a pattern
of deviations in flavor machines is likely to solve the new physics
flavor puzzle, to indirectly probe physics at a scale much higher than
the TeV scale, and, perhaps to make progress in understanding the
standard model flavor puzzle.

{\bf Acknowledgments }
We thank Kaustubh Agashe and Zoltan Ligeti  for comments on the first version of this review.
G.I. is supported by the EU under contract
MTRN-CT-2006-035482 {\em Flavianet}.
Y.N. is supported by the Israel Science Foundation (ISF) under grant
No.~377/07, by the German-Israeli foundation for scientific research
and development (GIF), and by the United States-Israel Binational
Science Foundation (BSF), Jerusalem, Israel. G.P. is supported by the Israel Science Foundation (grant \#1087/09), EU-FP7 Marie Curie, IRG fellowship and the Peter \&
Patricia Gruber Award.


\end{document}